\documentclass[prc,twocolumn,showpacs,superscriptaddress,preprintnumbers,amsmath,amssymb,nofootinbib]{revtex4-1}
\usepackage{bm}
\usepackage{dcolumn}
\usepackage{multirow}
\usepackage{ifpdf}
\usepackage{url}
\usepackage{comment}
    
\ifpdf
\usepackage{graphicx} 
\usepackage{hyperref}
\else 
\usepackage[dvipdfmx]{graphicx} 
\usepackage[dvipdfmx]{hyperref}
\fi
\usepackage{color} 
\usepackage{slashed}
\usepackage{bm}
\usepackage{dcolumn}
\usepackage{multirow}
\usepackage{ifpdf}
\usepackage{url}
\usepackage{here}
\usepackage{graphicx}
\usepackage{indentfirst}
\usepackage{listings}
\usepackage{ifthen}
\usepackage{kpfonts}
\usepackage{tgtermes}
\usepackage{amsmath,amssymb}
\usepackage{fancyhdr}
\usepackage{titletoc}
\usepackage{xcolor}
\usepackage{mathtools}

\makeatletter
\let\MYcaption\@makecaption
\makeatother
\usepackage{subcaption}
\makeatletter
\let\@makecaption\MYcaption
\makeatother

\newcommand{\dndeta}[1][flat]
{
    \ifthenelse{\equal{#1}{flat}}{$\left< dN_{\mathrm{ch}}/d\eta\right>$}{}
    \ifthenelse{\equal{#1}{vertical}}{$\left< \cfrac{dN_{\mathrm{ch}}}{d\eta\right>}$}{}
}
\newcommand{\snn}[1][nucleus]
{
    \ifthenelse{\equal{#1}{nucleus}}{$\sqrt{s_{\mathrm{NN}}}$}{}
    \ifthenelse{\equal{#1}{proton}}{$\sqrt{s}$}{}
}

\hypersetup{
colorlinks=true,
linkcolor=blue,
citecolor=blue,
urlcolor=blue
}

\begin{document}

\title{Causal hydrodynamic fluctuations in a one-dimensional expanding system}
\author{Shin-ei Fujii}
\email{s-fujii-8c2@eagle.sophia.ac.jp}
\affiliation{Department of Physics, Sophia University, Tokyo 102-8554, Japan}

\author{Tetsufumi Hirano}
\email{hirano@sophia.ac.jp}
\affiliation{Department of Physics, Sophia University, Tokyo 102-8554, Japan}

\date{\today}

\begin{abstract}
We derive stochastic equations of motion of hydrodynamic fluctuations, performing perturbative expansion of the energy-momentum conservation equations around the boost invariant solution in a one-dimensional expanding system.
In the course of derivation, we do not assume any specific forms of constitutive equations for shear stress tensor $\pi^{\mu \nu}$ and bulk pressure $\Pi$.
Therefore, the framework enables us to employ any constitutive equations beyond the Navier-Stokes theory which satisfy the causality.
Employing Israel-Stewart equations as examples of the constitutive equations, we demonstrate the dynamics of causal hydrodynamic fluctuations in (1+1)-dimensional Milne coordinates on an event-by-event basis.
We observe that the structure of energy density and flow rapidity fluctuations is almost frozen in the early stage of the expansion.
Two-point correlations of energy density fluctuations turn out to be closely related to the properties of the medium, such as sound velocity, viscosity, and relaxation time.
Furthermore, we show that two-particle correlation functions of final hadrons after freeze-out inherit correlations of thermodynamic variables and flow rapidity.
This opens a new door for an analysis of transport properties of the medium produced in relativistic heavy ion collisions. 
\end{abstract}

\maketitle
\section{Introduction}
One of the main goals of physics of relativistic heavy ion collisions is to reveal the bulk and the transport properties of the quark gluon plasma (QGP), namely, matter composed of quarks and gluons as elementary degrees of freedom under extremely high temperature and/or density. 
So far, a huge amount of experimental data on relativistic heavy ion collisions at the Relativistic Heavy Ion Collider (RHIC) and the Large Hadron Collider (LHC) has been reported.
Relativistic ideal hydrodynamic models  \cite{Kolb:2000fha, Huovinen:2001cy, Heinz:2001xi, Teaney:2000cw, Hirano:2001eu, Hirano:2002ds, Hirano:2005wx} greatly succeeded in describing the elliptic flow data at RHIC energies. 
Since then, relativistic hydrodynamics became a major framework to describe the space-time evolution of the QGP.
We are now in the stage to sophisticate the dynamical model and to use it to extract the properties of the QGP more precisely. 

Nowadays, relativistic dissipative hydrodynamics including shear and/or bulk viscosities has been used to extract transport properties of the QGP from experimental data through Bayesian parameter estimation \cite{Bernhard:2016tnd,Bernhard:2019bmu,Auvinen:2020mpc,Nijs:2020ors,Nijs:2020roc,Parkkila:2021tqq,JETSCAPE:2020mzn,Heffernan:2023gye,Shen:2023awv}.
Fluctuations and dissipations, however, always accompany each other according to the fluctuation-dissipation relations (FDR) in nonequilibrium statistical physics \cite{Kubo:1957mj}.
Since phenomena induced by the hydrodynamic fluctuations, namely, thermal fluctuations associated with the viscosities during hydrodynamic evolution, include the information of transport coefficients through FDRs, these provide a multidimensional analysis of transport properties of the QGP.
Therefore, it is indispensable to incorporate hydrodynamic fluctuations in the dynamical framework of relativistic heavy ion collisions.
This could open up a new way of diagnosing the QGP properties precisely.

In this paper, we formulate dynamics of causal hydrodynamic fluctuations in a one-dimensional expanding system \cite{Bjorken:1982qr} linearizing equations of energy-momentum conservation around the boost invariant solutions up to the first order without any assumption of specific constitutive equations.
We also linearize causal constitutive equations including noise terms around the boost invariant solutions. We regard them as stochastic differential equations for hydrodynamic fluctuations.
We demonstrate the event-by-event space-time evolution of fluctuations of thermodynamic variables and analyze the two-point correlation functions for them.
To see the effects of hydrodynamic fluctuations in observables, we also analyze two-particle correlation functions of final hadrons.

Linearization of hydrodynamic equations under a boost invariant solution \cite{Bjorken:1982qr} has been performed in, e.g., Refs.~\cite{Kouno:1989ps, Denicol:2008ha} for analysis of stability and causality against perturbation.
The first application of hydrodynamic fluctuations to the phenomenological model of relativistic heavy ion collisions was made in Ref.~\cite{Kapusta_2012}.
They linearized the relativistic hydrodynamic equations under boost invariant solutions \cite{Bjorken:1982qr} as backgrounds and regarded the first-order perturbative equations as equations of motion for hydrodynamic fluctuations. 
With these solutions, they obtained correlations of pion yield fluctuations in the rapidity direction.
This study was further extended to (3+1)-dimensional expansion \cite{Yan:2015lfa} by employing Gubser flow \cite{Gubser:2010ze} as a background.
However, both of them employed the first-order constitutive equations in which the gradient of flow velocity is included up to the first order for shear stress tensor $\pi^{\mu\nu}$ and bulk pressure $\Pi$.
It is well known that relativistic hydrodynamic equations with first-order theory are classified as parabolic equations: The time derivative is of the first order while the space derivative is of the second order.
Since the Green's function of parabolic equations is Gaussian and the long tail of it has a small but finite value, the propagation speed of sound exceeds the speed of light (acausal).
The first-order theory exhibits pathological behaviors and has not only an acausal problem but also instability \cite{Hiscock:1983zz, Hiscock:1985zz, Hiscock:1987zz, Kouno:1989ps, Denicol:2008ha}.
Therefore, one has to extend its  framework to the causal one
\cite{Israel:1976tn, Israel:1979wp, Hiscock:1983zz, Baier:2006um, Baier:2007ix, Betz:2008me, Betz:2009zz, Denicol:2012cn, Monnai:2010qp, Jaiswal:2013fc, Jaiswal:2013npa, Tsumura:2015fxa} in the stage of sophistication to study the properties of the QGP more precisely.

The first attempt to develop a framework of causal hydrodynamic fluctuations in a (1+1)-dimensional expanding system was done in Ref.~\cite{Chattopadhyay:2017rgh}, in which constitutive equations at the Navier-Stokes limit in Ref.~\cite{Kapusta_2012} were replaced with the second-order ones so that the system could obey the causality. 
Notice that ensemble averaged quantities were directly analyzed in Refs.~\cite{Kapusta_2012,Chattopadhyay:2017rgh}.
Here we concentrate more on the analysis of event-by-event phenomena induced by hydrodynamic fluctuations and of transport properties of the QGP from spectra of particle identified hadrons as an event average.
Instead of using the Green's function for the ensemble averaged quantities, we solve
system of stochastic partial differential equations for each event.
Furthermore, since hydrodynamics is an effective theory of the long wavelength limit, noise terms should be treated carefully to avoid singularities originated from short wavelengths.
We tame such a nonphysical singularity, seen in Refs.~\cite{Kapusta_2012,Chattopadhyay:2017rgh}, introducing a momentum cutoff for fluctuations in each event \cite{Murase:2015oie}.

There have already been also several studies based on (3+1)-dimensional hydrodynamic models which include causal hydrodynamic fluctuations \cite{Young:2013fka, Murase:2013tma, Murase:2015oie, Murase:2016rhl, Singh:2021mks, Sakai:2020pjw, Sakai:2021rug, Kuroki:2023ebq}.
It is, however, quite difficult to intuitively understand what results from hydrodynamic fluctuations in heavy-ion collisions due to their complicated dynamics. 
Hence a relatively simpler geometry like (1+1)-dimensional boost invariance is particularly useful to understand the consequences of hydrodynamic fluctuations even if both viscosity and fluctuations are incorporated perturbatively \cite{Kapusta_2012,Chattopadhyay:2017rgh}.
Thus, a study of causal hydrodynamic fluctuations in a one-dimensional expanding system has a practical advantage over the full (3+1)-dimensional hydrodynamic simulations thanks to a much simpler framework.
Since the hydrodynamic fluctuations originate from (local) thermal equilibrium, novel phenomena associated with hydrodynamic fluctuations would be clear and direct evidence of thermalization in relativistic heavy ion collisions.  

The present paper is organized as follows:
We derive a system of partial differential equations to describe the dynamics of causal hydrodynamic fluctuations in Sec.~\ref{sec:model}.
In Sec.~\ref{sec:results},  we first discuss the validity of perturbation and then analyze the space-time evolution of thermodynamic variables and their correlations.
We also calculate the two-particle correlation function of hadrons after freeze-out to see how the properties of the QGP affect observables.
Section \ref{sec:summary} is devoted to a summary of the present study.

Throughout this paper, we use natural units, $\hbar = c = k_{B} = 1$, and the Minkowski metric, $g_{\mu\nu} = \mathrm{diag}(1,-1,-1,-1)$.

\section{Model}
\label{sec:model}
In this section, we derive equations of motion (EoMs) of causal hydrodynamic fluctuations in a (1+1)-dimensional expanding system.
In Sec.~\ref{sec:balance equations}, we first derive the equations of background and of perturbation in a one-dimensional expanding system from the balance equations in relativistic hydrodynamics.
We also perform perturbative expansion for constitutive equations with hydrodynamic fluctuations employing Israel-Stewart equations \cite{Israel:1976tn, Israel:1979wp} in Sec.~\ref{sec:constitutive equations}.
We introduce fluctuation-dissipation relations (FDRs) in the Milne coordinate in Sec.~\ref{sec:FDR}.
We discuss some other variables in the perturbative expansion in Sec.~\ref{sec:other variables}.
We finally introduce models of the equation of state (EoS) and transport coefficients in Sec.~\ref{sec:model EoS}.
The formalism developed in this section is basically equivalent to the one in Ref.~\cite{Chattopadhyay:2017rgh} besides event-by-event treatment of dynamics including momentum cutoff in the noises in our study.

\subsection{Balance equations}
\label{sec:balance equations}
Hydrodynamic balance equations are composed of conservation laws of energy, momentum, and charges.
Throughout this paper, we neglect the conserved charges for simplicity. 
By means of tensor decomposition, the energy-momentum tensor, $T^{\mu\nu}$, can be expressed as
\begin{equation}\label{eq:E-Mtensor}
T^{\mu\nu} = eu^{\mu}u^{\nu} - (p+\Pi)\Delta^{\mu\nu} + \pi^{\mu\nu},
\end{equation}
where $e$, $p$, $\Pi$, and $\pi^{\mu\nu}$ are energy density, hydrostatic pressure, bulk pressure, and shear stress tensor, respectively.
As a definition of flow velocity $u^{\mu}$, we employ the Landau frame which satisfies an eigenvalue equation, $T^{\mu}_{\enskip \nu}u^{\nu} = eu^{\mu}$.
A projection tensor, $\Delta^{\mu\nu} \equiv g^{\mu\nu} - u^{\mu}u^{\nu}$, maps a four-vector to a space perpendicular to $u^{\mu}$.

A boost invariant solution in (1+1)-dimensional space for relativistic hydrodynamic equations is written as \cite{Bjorken:1982qr}
\begin{equation}\label{eq:Bj solution}
u_{\mathrm{Bj}}^{\mu} = \frac{t}{\tau}\left(1, 0, 0, \frac{z}{t}\right) = (\cosh \eta_{s},0 ,0, \sinh \eta_{s}),
\end{equation}
where $\tau \equiv \sqrt{t^{2}-z^{2}}$ and $\eta_{s} \equiv \tanh^{-1} (z/t)$ are proper time and space-time rapidity, respectively. 
This solution exhibits the one-dimensional Hubble-like expansion along the collision axis ($z$ axis), intrinsically holds boost invariant property, and, as a result, thermodynamic variables do not depend on $\eta_{s}$.

In order to perform perturbative expansion around boost invariant solutions, we first assume small deviations of flow velocity.
Flow velocity with small flow rapidity, $\delta y(\tau,\eta_{s})$, is
\begin{align}
u^{\mu} &= \left( \cosh \left( \eta_{s} + \delta y(\tau,\eta_{s}) \right), 0, 0, \sinh \left(\eta_{s} + \delta y(\tau,\eta_{s})\right) \right)\nonumber\\
&\approx \left( \cosh \eta_{s}, 0, 0, \sinh \eta_{s} \right)+
\left( \sinh\eta_{s} , 0, 0, \cosh\eta_{s} \right)\delta y\nonumber\\
&\equiv u^{\mu}_{(0)}(\eta_{s}) + u^{\mu}_{(1)} (\tau, \eta_{s}),\label{eq:velocity}
\end{align}
where subscripts $(0)$ and $(1)$ denote the zeroth and the first order in perturbation, respectively.
Correspondingly, thermodynamic variables and dissipative currents in the energy-momentum tensor (\ref{eq:E-Mtensor}) can be expanded as 
\begin{align}
e &\approx e_{(0)}(\tau) + e_{(1)}(\tau,\eta_{s}),\\
p &\approx p_{(0)}(\tau) + p_{(1)}(\tau, \eta_{s}), \\
\Pi &\approx \Pi_{(0)}(\tau) + \Pi_{(1)}(\tau, \eta_{s}), \\
\pi^{\mu\nu} &\approx \pi^{\mu\nu}_{(0)}(\tau) + \pi^{\mu\nu}_{(1)}(\tau, \eta_{s}).
\end{align}
Here the variables with subscript $(0)$ are independent of $\eta_{s}$ because of the boost invariant property of backgrounds.
Consequently, some other quantities become
\begin{align}
\Delta^{\mu\nu} &= g^{\mu\nu} - u^{\mu}u^{\nu} \approx g^{\mu\nu} - u^{\mu}_{(0)}u^{\nu}_{(0)} - u^{\mu}_{(0)}u^{\nu}_{(1)} - u^{\mu}_{(1)}u^{\nu}_{(0)}\nonumber\\
&\equiv \Delta^{\mu\nu}_{(0)} + \Delta^{\mu\nu}_{(1)},\\
\mathcal{D} &= u^{\mu}\partial_{\mu} \approx \frac{\partial}{\partial \tau} + \frac{\delta y}{\tau} \frac{\partial}{\partial \eta_{s}} \equiv \mathcal{D}_{(0)} + \mathcal{D}_{(1)}, \\
\nabla^{\alpha} &= \Delta^{\alpha\mu}\partial_{\mu} \approx \partial^{\alpha} - u^{\alpha}_{(0)}\mathcal{D}_{(0)} - u^{\alpha}_{(1)}\mathcal{D}_{(0)} - u^{\alpha}_{(0)}\mathcal{D}_{(1)}\nonumber\\
&\equiv \nabla^{\alpha}_{(0)} + \nabla^{\alpha}_{(1)}, \\
\theta &= \partial_{\mu}u^{\mu} \approx \frac{1}{\tau} + \frac{1}{\tau} \frac{\partial}{\partial \eta_{s}}  \delta y\equiv \theta_{(0)} + \theta_{(1)}.
\end{align}
Here a partial derivative is decomposed as $\partial^\mu = u^\mu \mathcal{D} + \nabla^\mu$ and $\theta$ is an expansion scalar.
Substituting these variables and differential operators into balance equations of the energy and momentum, $\partial_\mu T^{\mu \nu}=0$, we derive the EoMs of backgrounds and of fluctuations in a one-dimensional expanding system.
Timelike and spacelike parts of the energy-momentum conservation are, respectively,
\begin{align}
\label{eq:time-like conservation law}
u_{\nu}\partial_{\mu}T^{\mu\nu} &= 0,\\
\label{eq:space-like conservation law}
\Delta_{\alpha\nu}\partial_{\mu}T^{\mu\nu} &= 0.
\end{align}
From Eq.~(\ref{eq:time-like conservation law}),
we obtain
\begin{equation}\label{eq:generalform_energy}
\mathcal{D}e + (e + p + \Pi)\theta - \pi_{\mu\nu}\partial^{\langle\mu}u^{\nu\rangle} = 0,
\end{equation}
where angle brackets in the third term stand for the following operation for any second-rank tensor $A^{\mu\nu}$:
\begin{align}
A^{\langle\mu\nu\rangle} &\equiv \Delta^{\mu\nu}_{\enskip \enskip \alpha\beta}A^{\alpha\beta},\nonumber\\
&= \left[ \frac{1}{2}\left(\Delta^{\mu}_{\enskip \alpha}\Delta^{\nu}_{\enskip \beta} + \Delta^{\nu}_{\enskip \alpha}\Delta^{\mu}_{\enskip\beta}\right)  - \frac{1}{3}\Delta^{\mu\nu}\Delta_{\alpha\beta} \right]A^{\alpha\beta}.
\end{align}
The resultant tensor becomes symmetric, traceless, and transverse to flow velocity.
Inserting thermodynamic variables and differential operators up to the first order in perturbation into Eq.~(\ref{eq:generalform_energy}),
we obtain
\begin{align}
&\mathrm{0th:}\quad \frac{d}{d\tau}e_{(0)} + \frac{1}{\tau}\left(e_{(0)} + p_{(0)} + \Pi_{(0)} - \pi_{(0)}\right) = 0\label{eq:0th time EoM},\\
&\mathrm{1st:}\quad \frac{\partial}{\partial \tau}e_{(1)} + \frac{1}{\tau}\left(e_{(1)} + p_{(1)} + \Pi_{(1)} - \pi_{(1)}\right)\nonumber\\
&\quad\qquad+ \frac{1}{\tau}\frac{\partial}{\partial \eta_{s}} \delta y \left(e_{(0)} + p_{(0)} + \Pi_{(0)} - \pi_{(0)}\right) = 0,\label{eq:1st time EoM}
\end{align}
where we separate equations order by order and define shear pressure $\pi$ as
\begin{equation}
\pi  \equiv \pi^{00} - \pi^{33} \approx \pi_{(0)}(\tau) + \pi_{(1)}(\tau,\eta_{s}).
\end{equation}
The zeroth-order equation (\ref{eq:0th time EoM}) describes the time evolution of background energy density. This equation is nothing but the Bjorken equation with viscosity.
The first and the second terms in the left-hand side of Eq.~(\ref{eq:1st time EoM}) share the same form as those in Eq.~(\ref{eq:0th time EoM}), while the third term appears as a consequence of linearization.
The additional third term contains a gradient of fluctuations of flow rapidity with respect to $\eta_{s}$ and describes longitudinal dynamics of fluctuations.
Thus, the system is no longer boost invariant.

From Eq.~(\ref{eq:space-like conservation law}), we derive EoMs of flow velocity:
\begin{equation}\label{eq:general apace EoM}
(e + p + \Pi)\mathcal{D}u^{\alpha} - \nabla^{\alpha}(p + \Pi) + {\Delta^{\alpha}}_{\nu}\partial_{\mu}\pi^{\mu\nu} = 0.
\end{equation}
Following the same prescription as above, we finally obtain,\footnote{Actually, we obtain two equations which correspond to $\alpha = 0$ and $3$ in Eq. (\ref{eq:general apace EoM}). Linear combination of them results in this equation.
}
\begin{align}
\mathrm{0th:}\quad &\frac{\partial}{\partial \eta_{s}}(p_{(0)} + \Pi_{(0)} - \pi_{(0)}) = 0,\label{eq:0th space EoM}\\
\mathrm{1st:}\quad &\frac{\partial}{\partial \tau} \delta y\left(e_{(0)} + p_{(0)} + \Pi_{(0)} - \pi_{(0)}\right)\nonumber\\
&+ \frac{2\delta y}{\tau}\left(e_{(0)} + p_{(0)} + \Pi_{(0)} - \pi_{(0)}\right)\nonumber\\
&+ \frac{1}{\tau}\frac{\partial}{\partial \eta_{s}}(p_{(1)} + \Pi_{(1)} - \pi_{(1)}) = 0.\label{eq:1st space EoM}
\end{align}
The zeroth-order equation (\ref{eq:0th space EoM}) gives a condition for the background pressure and dissipative currents to be boost invariant.
Therefore, this equation can be neglected as long as we implicitly assume a boost invariant property of backgrounds. 
As seen in Eq.~(\ref{eq:1st time EoM}), a gradient with respect to $\eta_{s}$ appears in the third term of Eq.~(\ref{eq:1st space EoM}): The flow rapidity fluctuations are induced by the spatial gradient of fluctuations of total pressure, $p_{(1)} + \Pi_{(1)} - \pi_{(1)}$.
Since we do not assume any specific forms of constitutive equations, these balance equations obtained above, Eqs.~(\ref{eq:0th time EoM})-(\ref{eq:1st space EoM}), are generic and are used for any models of phenomenological constitutive equations.  

\subsection{Constitutive equations with noises}
\label{sec:constitutive equations}
In this paper, we employ the simplest causal constitutive equations derived by Israel and Stewart \cite{Israel:1976tn, Israel:1979wp} for the purpose of demonstration of the current framework,\footnote{These equations are often called ``simplified Israel-Stewart equations."
It is, however, not a precise name to describe these equations since what they actually showed in their papers are indeed these equations.
``Israel-Stewart equations" in the recent literature were obtained first by Hiscock and Lindbrom \cite{Hiscock:1983zz}.}
\begin{align}
\pi^{\mu\nu} + \Delta^{\mu\nu}_{\enskip \enskip \alpha\beta}\tau_{\pi}\mathcal{D}\pi^{\alpha\beta} &= 2\eta\nabla^{\langle\mu}u^{\nu\rangle} + \xi_{\pi}^{\mu\nu},\label{eq:shear full}\\
(1 + \tau_{\Pi}\mathcal{D})\Pi &= -\zeta\theta + \xi_{\Pi},
\end{align}
where noise terms $\xi_{\pi}^{\mu\nu}(\tau,\eta_{s})$ and $\xi_{\Pi}(\tau,\eta_{s})$ for shear stress tensor and bulk pressure, respectively, are introduced as hydrodynamic fluctuations.
Here transport coefficients $\eta$, $\zeta$, $\tau_{\pi}$, and $\tau_{\Pi}$  are shear viscosity, bulk viscosity, relaxation time for shear stress tensor, and relaxation time for bulk pressure, respectively.
Since the transport coefficients are, in general, functions of temperature, which has boost invariant background and small fluctuations through the EoS, the transport coefficients can be also decomposed into the boost invariant zeroth-order terms and the space-time rapidity dependent first-order terms as fluctuations: 
\begin{align}
\eta &\approx \eta_{(0)}(\tau) + \eta_{(1)}(\tau, \eta_{s}), \\
\zeta &\approx \zeta_{(0)}(\tau) + \zeta_{(1)}(\tau, \eta_{s}),\\
\tau_{\pi} &\approx \tau_{\pi(0)}(\tau) + \tau_{\pi(1)}(\tau, \eta_{s}), \\
\tau_{\Pi} &\approx \tau_{\Pi(0)}(\tau) + \tau_{\Pi(1)}(\tau, \eta_{s}).
\end{align}

With these assumptions, we finally obtain the constitutive equations for shear pressure, $\pi$, as
\begin{align}
\mathrm{0th:}\quad & \pi_{(0)} + \tau_{\pi(0)}\frac{d}{d \tau} \pi_{(0)} = \frac{4\eta_{(0)}}{3\tau}\label{0th shear IS final}, \\
\mathrm{1st:}\quad & \pi_{(1)} + \tau_{\pi(0)} \frac{\partial}{\partial \tau} \pi_{(1)} + \tau_{\pi(1)} \frac{d}{d \tau} \pi_{(0)}\nonumber\\ 
&= \frac{4\eta_{(1)}}{3\tau} + \frac{4\eta_{(0)}}{3\tau}\frac{\partial}{\partial \eta_{s}} \delta y + \xi_{\pi}\label{1st shear IS final},
\end{align}
where, for both the zeroth-order and the first-order perturbative equations, shear stress tensor $\pi^{\mu\nu}$ in Eq.~(\ref{eq:shear full}) is simply reduced to $\pi_{(0)} = \pi_{(0)}^{00} - \pi_{(0)}^{33}$ and $\pi_{(1)} = \pi_{(1)}^{00} - \pi_{(1)}^{33}$ within the present assumptions. The noise term $\xi_{\pi}$ is also defined as $\xi_{\pi} \equiv \xi_{\pi}^{00}-\xi_{\pi}^{33}$.
Similarly, the constitutive equations for bulk pressure $\Pi$ are obtained as
\begin{align}
\mathrm{0th:}\quad & \Pi_{(0)} + \tau_{\Pi(0)}\frac{d}{d \tau} \Pi_{(0)} = -\frac{\zeta_{(0)}}{\tau},\label{0th bulk IS final}\\
\mathrm{1st:}\quad & \Pi_{(1)} + \tau_{\Pi(0)} \frac{\partial}{\partial \tau} \Pi_{(1)} + \tau_{\Pi(1)} \frac{d}{d \tau} \Pi_{(0)}\nonumber\\
&= -\frac{\zeta_{(1)}}{\tau} - \frac{\zeta_{(0)}}{\tau}\frac{\partial}{\partial \eta_{s}} \delta y + \xi_{\Pi}\label{1st bulk IS final}.
\end{align}

In the course of derivation, we regarded noise terms $\xi_{\pi}$ and $\xi_{\Pi}$ as variables at the first order \cite{Kapusta_2012,Chattopadhyay:2017rgh}.
In other words, fluctuations of thermodynamic variables on top of boost-invariant background are induced directly ($\pi_{(1)}$ and $\Pi_{(1)}$) or indirectly ($e_{(1)}$ and $\delta y$) by hydrodynamic fluctuations.

\subsection{Fluctuation-dissipation relation}
\label{sec:FDR}
We next set the power of noises and their probability distributions.
When the background medium keeps local equilibrium but dynamically evolves, as is often assumed in the space-time evolution in relativistic heavy ion collisions, it is not trivial whether the ordinary FDRs can be used.
The FDR was generalized in such a case in Ref.~\cite{Murase:2019cwc}.
Although we should have employed this generalized version of FDR in the current setting, we postpone analysis of the effect of generalization to future work and employ the ordinary FDR in this study.

From the consequences of nonequilibrium statistical physics, Gaussian white noises, $\xi_{\pi}$ and $\xi_{\Pi}$, obey the following FDRs in the Milne coordinate \cite{Hirano:2018diu}:
\begin{align}
\langle \xi_{\pi}(x) \rangle & = 0,\label{eq:Mean of noise shear}\\
\langle \xi_{\pi}(x)\xi_{\pi}(x^{\prime}) \rangle &= \frac{8\eta_{(0)} T_{(0)}}{3} \delta^{4}(x-x^{\prime})\nonumber\\
&= \frac{8\eta_{(0)} T_{(0)}}{3\tau} \delta(\tau-\tau^{\prime})\delta(\eta_{s}-\eta_{s}^{\prime})\delta^{2}(\bm{x}_{\perp}-\bm{x}_{\perp}^{\prime}), \label{eq:FDR for shear Bj}\\
\langle \xi_{\Pi}(x) \rangle & = 0,\label{eq:Mean of noise bulk}\\
\langle \xi_{\Pi}(x)\xi_{\Pi}(x^{\prime}) \rangle &= 2\zeta_{(0)} T_{(0)} \delta^{4}(x-x^{\prime})\nonumber\\
&= \frac{2\zeta_{(0)} T_{(0)}}{\tau} \delta(\tau-\tau^{\prime})\delta(\eta_{s}-\eta_{s}^{\prime})\delta^{2}(\bm{x}_{\perp}-\bm{x}_{\perp}^{\prime}),\label{eq:FDR for bulk Bj}
\end{align}
where $T_{(0)}$ is the temperature of the background and $\bm{x}_{\perp} \equiv (x, y)$ are transverse coordinates. 
Regarding the FDR of shear pressure (\ref{eq:FDR for shear Bj}), the original form has four Lorentz indices,
\begin{equation}
\langle \xi_{\pi}^{\mu\nu}(x)\xi_{\pi}^{\alpha\beta}(x^{\prime}) \rangle = 4\eta_{(0)}T_{(0)}\delta^{4}(x-x^{\prime})\Delta^{\mu\nu\alpha\beta}.\label{eq:FDR for shear Bj origin}
\end{equation}
Under the present situations, the noise term is, however, no longer a tensor, and is reduced to $\xi_{\pi}(x) \equiv \xi_{\pi}^{00}(x) - \xi_{\pi}^{33}(x)$ due to the symmetry. Thus, Eq.~(\ref{eq:FDR for shear Bj}) can be obtained from Eq.~(\ref{eq:FDR for shear Bj origin}) through the following calculations:
\begin{align}
&\langle \xi_{\pi}(x)\xi_{\pi}(x^{\prime}) \rangle \nonumber\\
&= \big\langle \left(\xi_{\pi}^{00}(x)-\xi_{\pi}^{33}(x)\right) \left(\xi_{\pi}^{00}(x^{\prime})-\xi_{\pi}^{33}(x^{\prime})\right) \big\rangle \nonumber\\
&= 4\eta_{(0)}T_{(0)}\delta^{4}(x-x^{\prime}) \left( \Delta^{0000} + \Delta^{3333} - \Delta^{0033} - \Delta^{3300}\right)\nonumber\\
&= \frac{8\eta_{(0)} T_{(0)}}{3} \delta^{4}(x-x^{\prime}).
\end{align}

When it comes to solving the stochastic differential equations numerically, both space and time should be discretized by introducing finite time step $\Delta \tau$ and cell size $\Delta \eta_{s}$. Regarding FDRs (\ref{eq:FDR for shear Bj}) and (\ref{eq:FDR for bulk Bj}), delta functions should be also discretized by the following replacement:
\begin{align}
\delta(\tau - \tau^{\prime}) &\rightarrow \frac{\delta_{\tau\tau^{\prime}}}{\Delta\tau},\\
\delta^{2}(x_{\perp} - x^{\prime}_{\perp}) &\rightarrow \frac{1}{\Delta x \Delta y},\label{eq:delta trans}\\
\delta(\eta_{s} - \eta_{s}^{\prime}) &\rightarrow G(\eta_{s} - \eta_{s}^{\prime}) \equiv \frac{1}{\sqrt{2\pi\sigma_{\eta_{s}}^2}} \text{exp}\left( -\frac{(\eta_{s} - \eta_{s}^{\prime})^{2}}{2\sigma_{\eta_{s}}^{2}} \right),
\end{align}
where the delta function of $\eta_{s}$ is replaced with a Gaussian function with finite standard deviation $\sigma_{\eta_{s}}$ so that spatially smeared noises are generated \cite{Murase:2015oie}.
To avoid singularities originating from short wavelengths of hydrodynamic fluctuations,
we introduce the momentum cutoff, $1/\sigma_{\eta_{s}}$.
This essentially suppresses the higher wave numbers (momenta) of noises in the space-time rapidity direction. 
For the time integration of stochastic differential equations, the second-order stochastic Runge-Kutta method is employed \cite{SDE, Hirano:2018diu}.

Throughout this paper, we set $\Delta \tau = 0.01$ fm, $\Delta x = \Delta y = 2$ fm, $\Delta\eta_{s} = 0.1$, and $\sigma_{\eta_{s}} = 0.1$.
We introduce the smearing of hydrodynamic noises with a cutoff parameter $(= 5\sigma_{\eta_{s}})$ and the noises are no longer correlated beyond the cutoff parameter length in the space-time rapidity direction.
The region of space-time rapidity is defined in $\eta_{s, \mathrm{min}} \le \eta_s \le \eta_{s, \mathrm{max}}$ with $\eta_{s, \mathrm{min}}=0$ and $\eta_{s, \mathrm{max}}=10$.
We employ periodic boundary conditions of the first-order variables, $e_{(1)} (\eta_{s} = \eta_{s, \mathrm{min}}) = e_{(1)} (\eta_{s} = \eta_{s, \mathrm{max}})$, and so on in the numerical calculations. 

\subsection{Other variables}
\label{sec:other variables}
In order to deal with the first-order variables of transport coefficients, we assume they depend on temperature and convert variables into fluctuations of energy density $e_{(1)}$. Then, the first-order variables of transport coefficients are written as
\begin{align}
\tau_{\pi(1)}(T) &= \frac{\partial \tau_{\pi(0)}(T)}{\partial T} \frac{\partial T}{\partial e_{(0)}} e_{(1)},\\
\tau_{\Pi(1)}(T) &= \frac{\partial \tau_{\Pi(0)}(T)}{\partial T} \frac{\partial T}{\partial e_{(0)}} e_{(1)},\\
\eta_{(1)}(T) &= \frac{\partial \eta_{(0)}(T)}{\partial T} \frac{\partial T}{\partial e_{(0)}} e_{(1)},\\
\zeta_{(1)}(T) &= \frac{\partial \zeta_{(0)}(T)}{\partial T} \frac{\partial T}{\partial e_{(0)}} e_{(1)}.
\end{align}
Here $\partial e_{(0)}/\partial T \equiv c_{V}(T)$ is heat capacity per unit volume. 
Once specific models of transport coefficients are given, the first-order variables can be treated in this way.

The first-order term of hydrostatic pressure is obtained after a specific model of the EoS, $p=p(e)$, is given as
\begin{equation}
p_{(1)} = \frac{\partial p_{(0)}}{\partial e_{(0)}} e_{(1)} = c_{s}^{2} e_{(1)},
\end{equation}
where $c_{s}^{2}$ is the square of sound velocity.

\subsection{Models of EoS and transport coefficients}
\label{sec:model EoS}
In the following, we specify models of EoS and transport coefficients.
We employ two models of the EoS: the conformal EoS, $p = \frac{1}{3}e$, with the degrees of freedom $d = 47.5$, for massless $N_f = 3$ QCD as one model, and a parametrization of lattice EoS results \cite{HotQCD:2014kol} as the other model.
In the case of conformal EoS, bulk pressure $\Pi$ vanishes because of conformal symmetry.
Hence we also neglect the bulk pressure $\Pi$ even in the case of lattice EoS for comparison and focus on the shear pressure $\pi$ as a dissipative current.
For the transport coefficients, we choose the specific shear viscosity $\eta/s = 1/4\pi$ \cite{Kovtun:2004de} and relaxation time $\tau_{\pi} = (2-\text{ln}2)/2\pi T$ \cite{Baier:2007ix} obtained from AdS/CFT correspondence.
These parameters are used as a default setting.
When we investigate parameter dependence on final results, we multiply these transport coefficients by a constant factor.

\section{Results}
\label{sec:results}
In what follows, we change the notation of background energy density as $e_{(0)} \rightarrow e_{0}$ and a fluctuation of energy density as $e_{(1)} \rightarrow \delta e$, for simplicity.
Then, the total energy density of the system is written as
$e(\tau, \eta_{s}) = e_{0}(\tau) + \delta e(\tau, \eta_{s})$.
The notation of the other variables is also changed accordingly. 
Since the equations to be solved are first-order differential equations in time, we need to assign initial conditions for each variable.
We start the hydrodynamic evolution at initial time $\tau_{\mathrm{ini}} = 1 \;\text{fm}$.
Initial conditions are summarized in Table \ref{tb:InitialCond}.
These initial conditions are commonly used throughout this paper.\footnote{When we study effects of transport coefficients on final results by multiplying a factor with the shear viscosity $\eta_0$ in Figs.~\ref{fig:shear1} and \ref{fig:particle_correlation_shear}, the initial conditions for $\pi_0$ are also changed accordingly.}
The smooth initial conditions with vanishing fluctuations are chosen in the longitudinal direction.
We plan to investigate the effects of initial longitudinal fluctuations on final observables in future publications.

\begin{table}[htbp]
\centering
\caption{Initial conditions.}
\begin{tabular}{ll} \hline \hline 
Variables \hspace{30pt} & Values \\ \hline
$\tau$ & 1  ${\rm fm}$\\
$e_{0}$ & 10  ${\rm GeV/fm^{3}}$\\
$\pi_{0}$ & $4\eta_{0} / 3\tau_{\mathrm{ini}}$  ${\rm GeV/fm^{3}}$\\
$\delta e$ & 0  ${\rm GeV/fm^{3}}$\\
$\delta \pi$ & 0  ${\rm GeV/fm^{3}}$\\
$\delta y$ & 0  \\\hline
\end{tabular}
\label{tb:InitialCond}
\end{table}

Before going into details on the results, we discuss how the sound wave propagates in a one-dimensional expanding system in Appendix \ref{sec:A}.
The property of sound propagation in a one-dimensional expanding system is totally different from that of a static medium.
When the reference of the frame moves at some constant speed, one observes the sound horizon.
On the other hand, when the background medium expands, one sees that the information of fluctuations reaches infinity in $\eta_{s}$ space.
These are helpful in understanding how the individual fluctuation induced by thermal noises propagates in the space-time rapidity direction. 

\subsection{Validity of perturbation}
\label{sec:validity}

\begin{figure}[tbp]
    \begin{minipage}[b]{1.0\linewidth}
    \centering
    \includegraphics[clip,width=\linewidth]{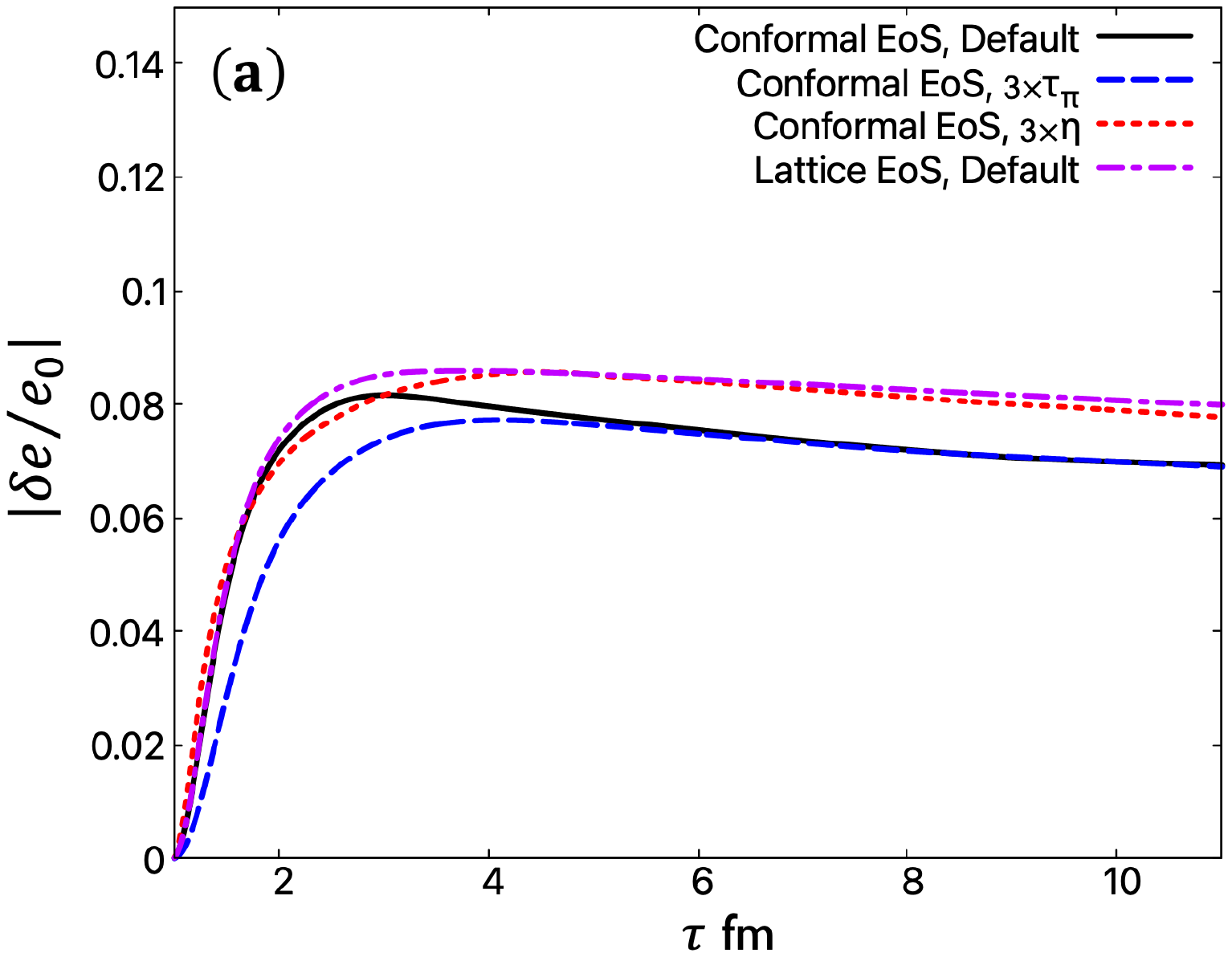}
    \end{minipage}\\
    \begin{minipage}[b]{1.0\linewidth}
    \centering
    \includegraphics[clip,width=\linewidth]{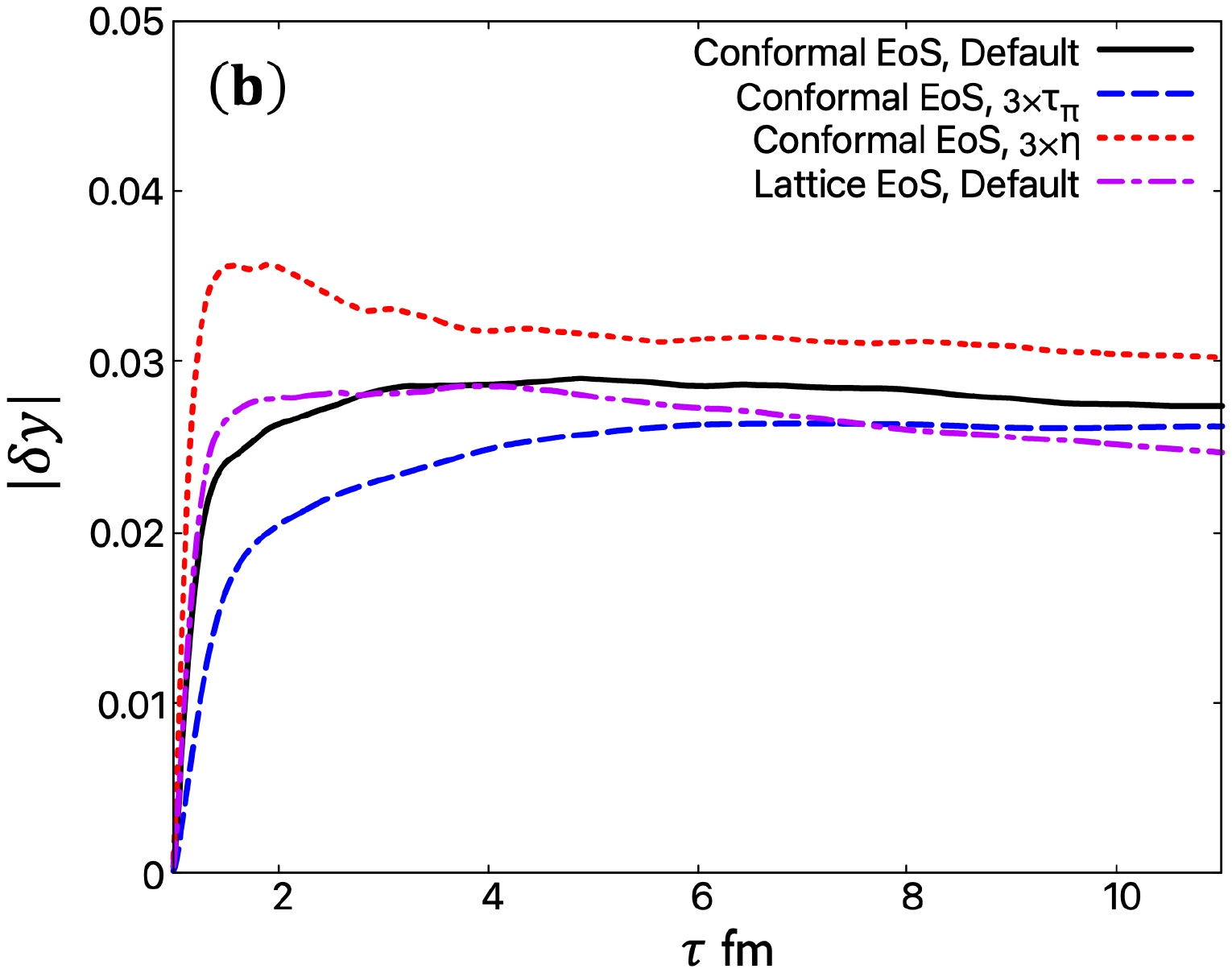}
    \end{minipage}\\
    \begin{minipage}[b]{1.0\linewidth}
    \centering
    \includegraphics[clip,width=\linewidth]{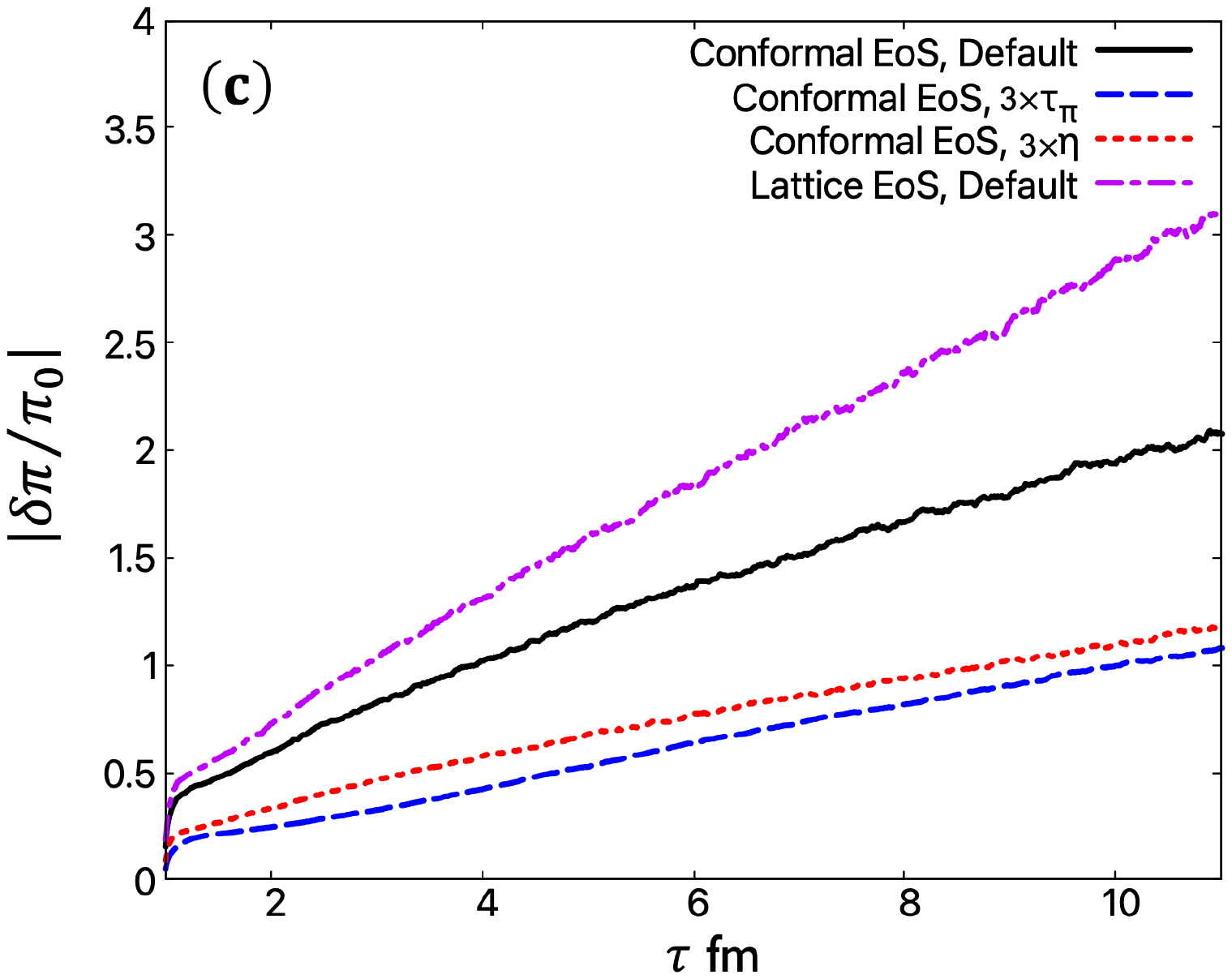}
    \end{minipage}
    \caption{Absolute values of the ratio of average fluctuations and background. Time evolution of (a) energy density, (b) flow rapidity, and (c) shear pressure are shown in comparison.
    The black solid line, blue dashed line, red dotted line, and magenta dash-dotted line are results from conformal EoS with default transport coefficients, conformal EoS with three times larger $\tau_{\pi}$, conformal EoS with three times larger $\eta$, and lattice EoS with default transport coefficients, respectively.
    All results are averaged over 10,000 events at $\eta_{s} = 5.0$.}
    \label{fig:ratio}
\end{figure}

First of all, we discuss the validity of perturbation and clarify the applicability of our model.
Since we linearized the EoMs under the assumption that fluctuations are sufficiently smaller than backgrounds, we need to care about the magnitude of fluctuations.
Figure \ref{fig:ratio} shows the time evolution of the absolute values of the ratios (a) $|\delta e/e_{0}|$, (b) $|\delta y|$, and (c) $|\delta \pi/\pi_{0}|$.
Here the magnitude of fluctuation of flow rapidity, $\delta y$, should be compared to just unity since flow velocity (\ref{eq:velocity}) in the comoving system ($y=\eta_{s}$) is
\begin{equation}
u_{\mathrm{comoving}}^{\mu} = (1, 0, 0, \delta y).
\end{equation}

It is clear that the ratios are always smaller than $0.1$ for energy density and flow rapidity, while the ratio of shear pressure monotonically increases and exceeds unity at $\tau \approx 4\;\text{fm}$ in the case of conformal EoS and default transport coefficients.
On the other hand, in the case of lattice EoS, it exceeds unity at $\tau \approx 2.5$ fm.
To understand the EoS dependence of the behaviors of $|\delta \pi/\pi_{0}|$ in Fig.~\ref{fig:ratio} (c), let us analytically assess the time dependence of $|\xi_{\pi}|/\pi_0$ instead of $\delta\pi/\pi_0$ by assuming a model EoS, $p=c_s^2 e$, and shear viscosity, $\eta/s = 1/4\pi$, at the first order theory.
First, time dependence of $\pi_{0}$ can be estimated as
\begin{equation}
\pi_{0} = \frac{4\eta_{0}}{3\tau} \propto s_{0}\times \tau^{-1} \propto \tau^{-2},\label{eq:time dependence background}
\end{equation}
where $s_{0}$ is the entropy density of the background and the effect of entropy production on time dependence can be neglected. 
On the other hand, time dependence of the standard deviation of noise $\xi_{\pi}$ can be given by the FDR (\ref{eq:FDR for shear Bj}),
\begin{equation}
|\xi_{\pi}| \approx \langle \xi_{\pi}\xi_{\pi} \rangle^{\frac{1}{2}} \propto \left(\frac{\eta_{0}T_{0}}{\tau}\right)^{\frac{1}{2}} \propto\left( s_{0} T_{0} \tau^{-1}\right)^{\frac{1}{2}} \propto \tau^{-1-\frac{1}{2}c_s^2}.\label{eq:time dependence fluctuation}
\end{equation}
Therefore, time dependence of a typical value of the ratio $|\xi_{\pi}|/\pi_{0}$ leads to
\begin{equation}
\frac{|\xi_{\pi}|}{\pi_{0}} \propto \tau^{1-\frac{1}{2}c_s^2}.\label{eq:time dependence of noise ratio}
\end{equation}
Since the power of $\tau$ in Eq.~(\ref{eq:time dependence of noise ratio}) is always positive, $1-\frac{1}{2}c_s^2\ge 0$, the ratio monotonically increases with proper time, and fluctuations become dominant against the background shear pressure eventually.\footnote{The actual power would be corrected due to the production of entropy. Nevertheless, the conclusion does not change here.}
Moreover, the softer the EoS is, the more rapidly the ratio increases.
This is the reason why the ratio with lattice EoS increases more rapidly than the one with conformal EoS.
When it becomes larger than unity, this framework breaks down.

The stability condition of the thermal state is nothing but the FDR. 
Thus, in the thermal bath, the background shear stress tensor vanishes on average, while fluctuations of it always exist due to thermal fluctuations. 
Since the expansion rate of the volume decreases as $\theta \approx \tau^{-1}$ in a one-dimensional expanding system, it is inevitable that the fluctuations of shear pressure becomes larger than its background and that the current framework eventually breaks down.

As a possible makeshift prescription to this issue, we may use larger values of $\Delta x$ and $\Delta y$ in Eq.~(\ref{eq:delta trans}) to suppress the noise, or use larger initial conditions for background.
As seen in Fig.~\ref{fig:ratio}, the larger the chosen transport coefficients are, the better the perturbative treatment is.
This can be understood from Eqs.~(\ref{eq:time dependence background}) and (\ref{eq:time dependence fluctuation}) since the ratio becomes
\begin{equation}
\frac{|\xi_{\pi}|}{\pi_{0}} \propto \frac{1}{\sqrt{\eta_0}}
\end{equation}
at a fixed time.

Although we should have had to pay attention to the behavior of shear pressure in the present study, we postpone a detailed analysis of conditions for the validity of perturbation to a future publication, which includes reformulation of equation of motion for shear pressure.
In Sec.~\ref{sec:two-particle correlation}, we will boldly use the information in the later stage under hydrodynamic evolution with lattice EoS when we calculate two-particle correlation functions, which might not have been justified from a viewpoint of validity conditions discussed in this section.

\subsection{Space-time evolution}
\begin{figure}[tbp]
    \centering
    \includegraphics[clip,width=1\linewidth]{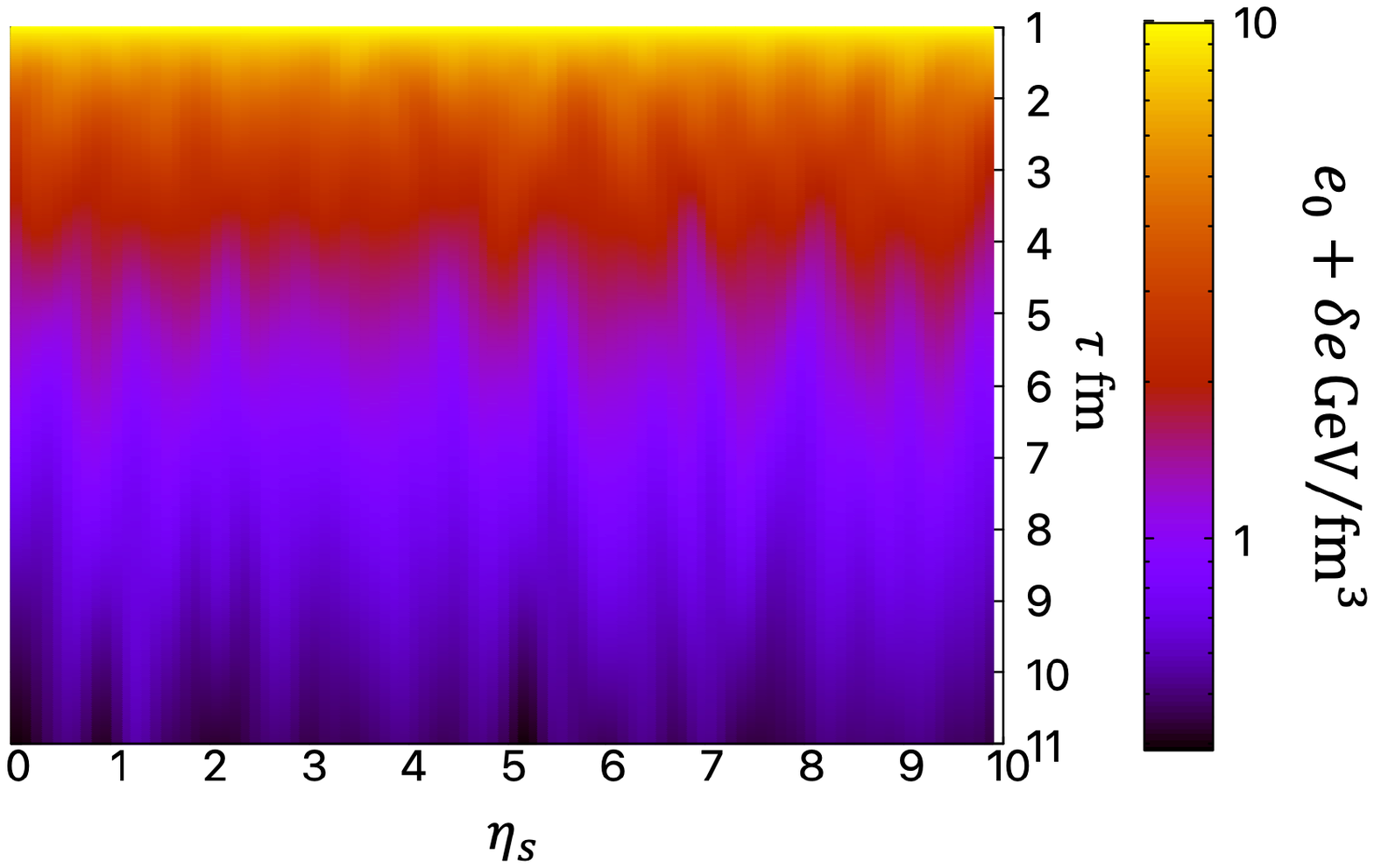}
    \caption{Space-time evolution of total energy density in an event as an example. Time passes from top to bottom and the color bar indicates the magnitude of energy density.}
    \label{fig:space-time_evolution_e}
\end{figure}

Next, let us exhibit numerical solutions of a system of stochastic partial differential equations obtained in the previous section.
First, we describe the space-time evolution of thermodynamic variables, which is a sum of the background and fluctuations. 
Figure \ref{fig:space-time_evolution_e} shows the time evolution of energy density distribution from one sampled event.\footnote{The tendencies of time evolution exhibited in this subsection are quite common in other sampled events.}
It is evident that the background energy density decreases immediately from the initial value, $e_{0}(\tau = 1\;\text{fm}) = 10\;{\rm GeV/fm^{3}}$, due to the rapid expansion of the system. Through a system of partial differential equations, the fluctuations of energy density are induced by hydrodynamic fluctuations: Hydrodynamic fluctuations induce the fluctuations of shear pressure, $\delta \pi$, which gives a space-time rapidity dependent $pdV$ work; on the other hand, fluctuations of flow rapidity, $\delta y$, also give a space-time dependent expansion rate of local volume, $\delta \theta$.
A crucial thing is that a streak-like structure appears through the time evolution and is kept until the final time $\tau = 11\;\text{fm}$.
It means that the pattern of energy density distribution is almost frozen during the evolution and could carry the information of the early stage.
One of the possible reasons for such a phenomenon is the interplay between the diffusion of fluctuations due to the finite shear viscosity and the effect of stretching the fluctuations due to the rapid expansion of the system. 
That is to say, the phenomenon ``freeze of distribution'' is an intrinsic property of an expanding system.
To our best knowledge, it is shown for the first time that hydrodynamic fluctuations lead to structure formation of matter created in relativistic heavy ion collisions.

\begin{figure}[tbp]
    \centering
    \includegraphics[clip,width=1\linewidth]{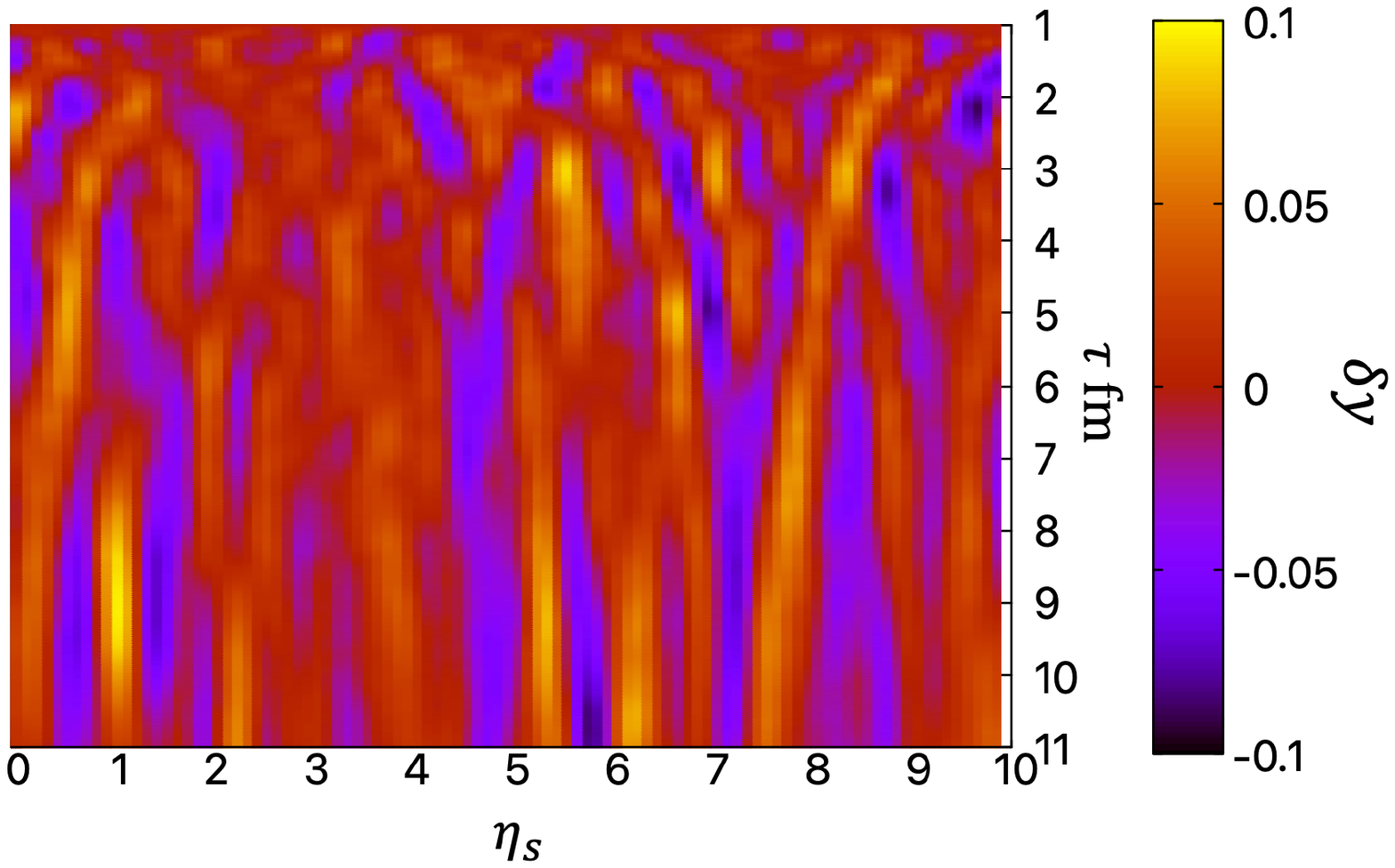}
    \caption{Space-time evolution of flow rapidity fluctuations in the same event as in Fig.~\ref{fig:space-time_evolution_e}. Time passes from top to bottom and the color bar indicates the magnitude of flow rapidity fluctuations.}
    \label{fig:space-time_evolution_y}
\end{figure}

In Fig.~\ref{fig:space-time_evolution_y}, a similar structure appears in the case of flow rapidity fluctuations. 
From Eq.~(\ref{eq:1st space EoM}), it is induced by the spatial gradient of fluctuations of energy density, $\delta e$, and those of shear pressure $\delta \pi$. 
The gradient of energy density fluctuations persists throughout the time evolution.
Note that we plotted only flow rapidity fluctuations, $\delta y$, rather than the total flow rapidity in Fig.~\ref{fig:space-time_evolution_y} since we are interested in deviation from  Bjorken's solution (\ref{eq:Bj solution}).

\begin{figure}[tbp]
    \centering
    \includegraphics[clip,width=1\linewidth]{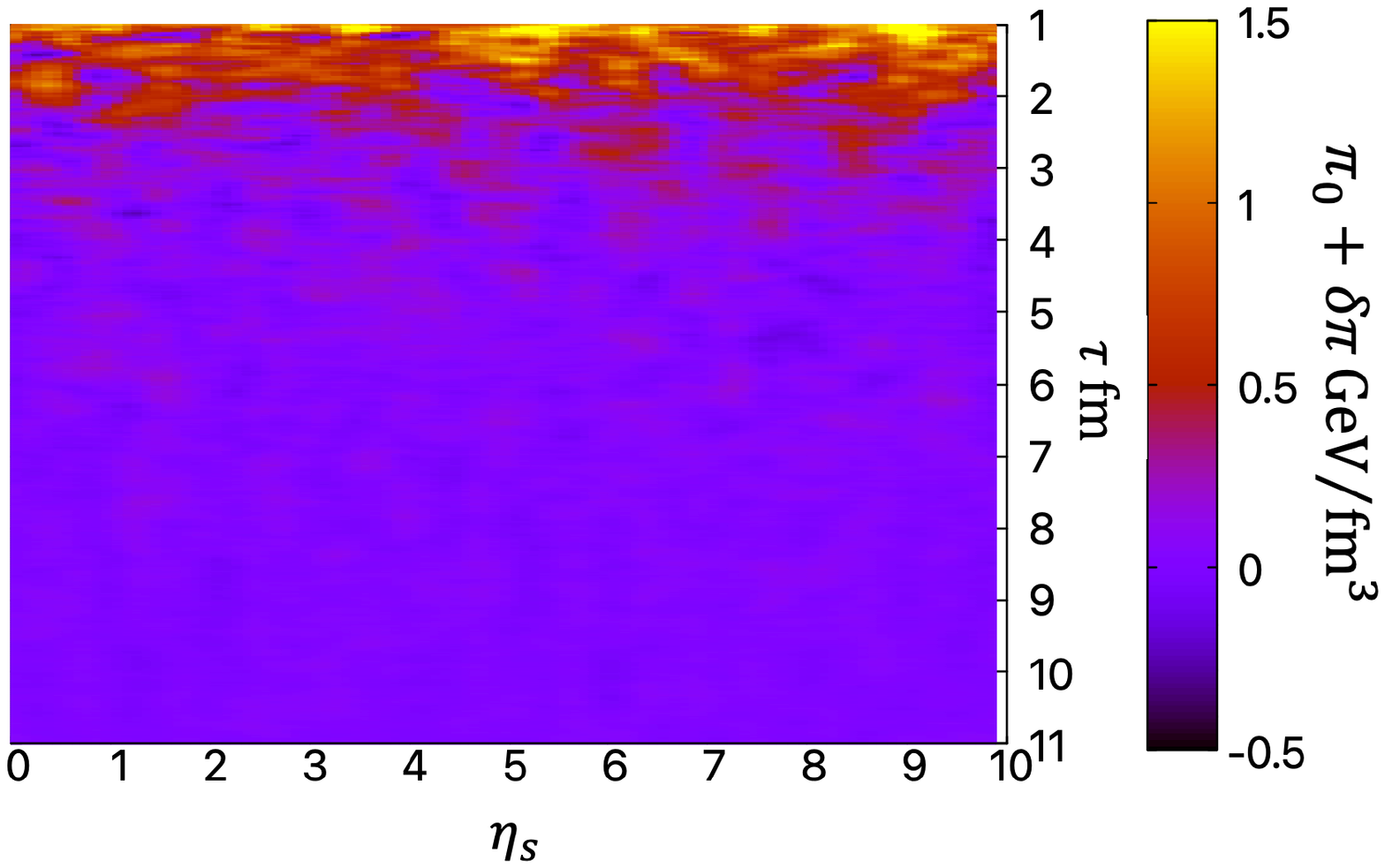}
    \caption{Space-time evolution of total shear pressure in the same event as in Fig.~\ref{fig:space-time_evolution_e}. Time passes from top to bottom and the color bar indicates the magnitude of shear pressure.}
    \label{fig:space-time_evolution_pi}
\end{figure}

Figure \ref{fig:space-time_evolution_pi} shows the space-time evolution of shear pressure.
In contrast to the case of energy density or flow rapidity, there is no such a streak-like structure in the shear pressure.
Shear pressure is known as a ``fast variable,'' namely, fluctuations of shear pressure are damped very quickly since it is not a conserved variable.
This is one of the main reasons why the streak-like structure cannot be observed in the space-time evolution of shear pressure. 

\subsection{Correlations of fluctuations}
\begin{figure}[tbp]
    \begin{minipage}[b]{1.0\linewidth}
    \centering
    \includegraphics[clip,width=\linewidth]{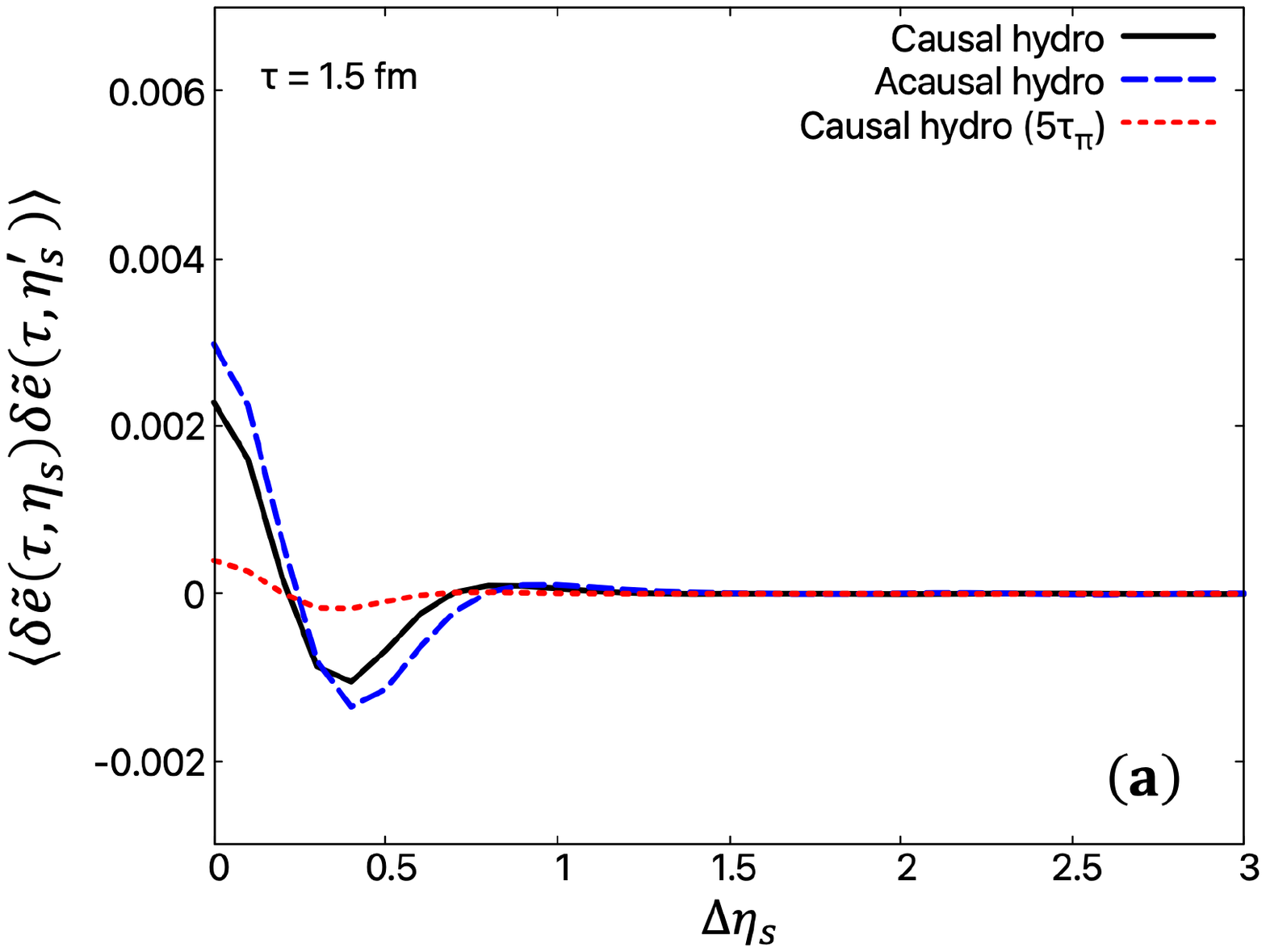}
    \end{minipage}\\
    \begin{minipage}[b]{1.0\linewidth}
    \includegraphics[clip,width=\linewidth]{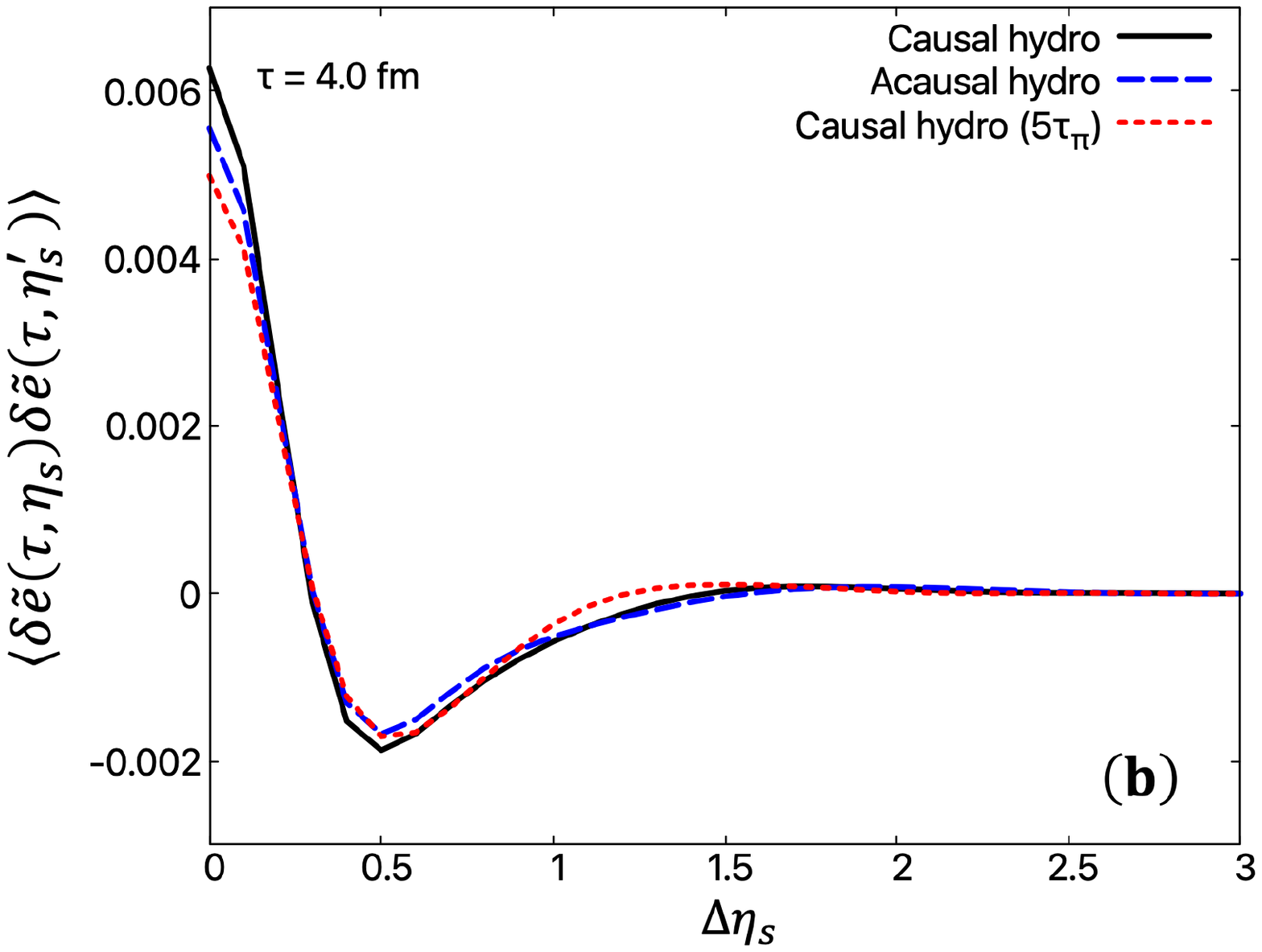}
    \end{minipage}
    \caption{Normalized correlations of energy density fluctuations as functions of $\Delta \eta_{s} \equiv |\eta_{s} - \eta_{s}^{\prime}|$.
    We make a comparison between causal hydro (default settings, black solid line), acausal hydro ($\tau_{\pi}=0\;\text{fm}$, blue dashed line), and causal hydro ($5\tau_{\pi}$, red dotted line) at (a) $\tau = 1.5 \;\text{fm}$ and (b) $\tau = 4.0 \;\text{fm}$.
     }
    \label{fig:order1}
\end{figure}

\begin{figure}[tbp]
    \begin{minipage}[b]{1.0\linewidth}
    \centering
    \includegraphics[clip,width=\linewidth]{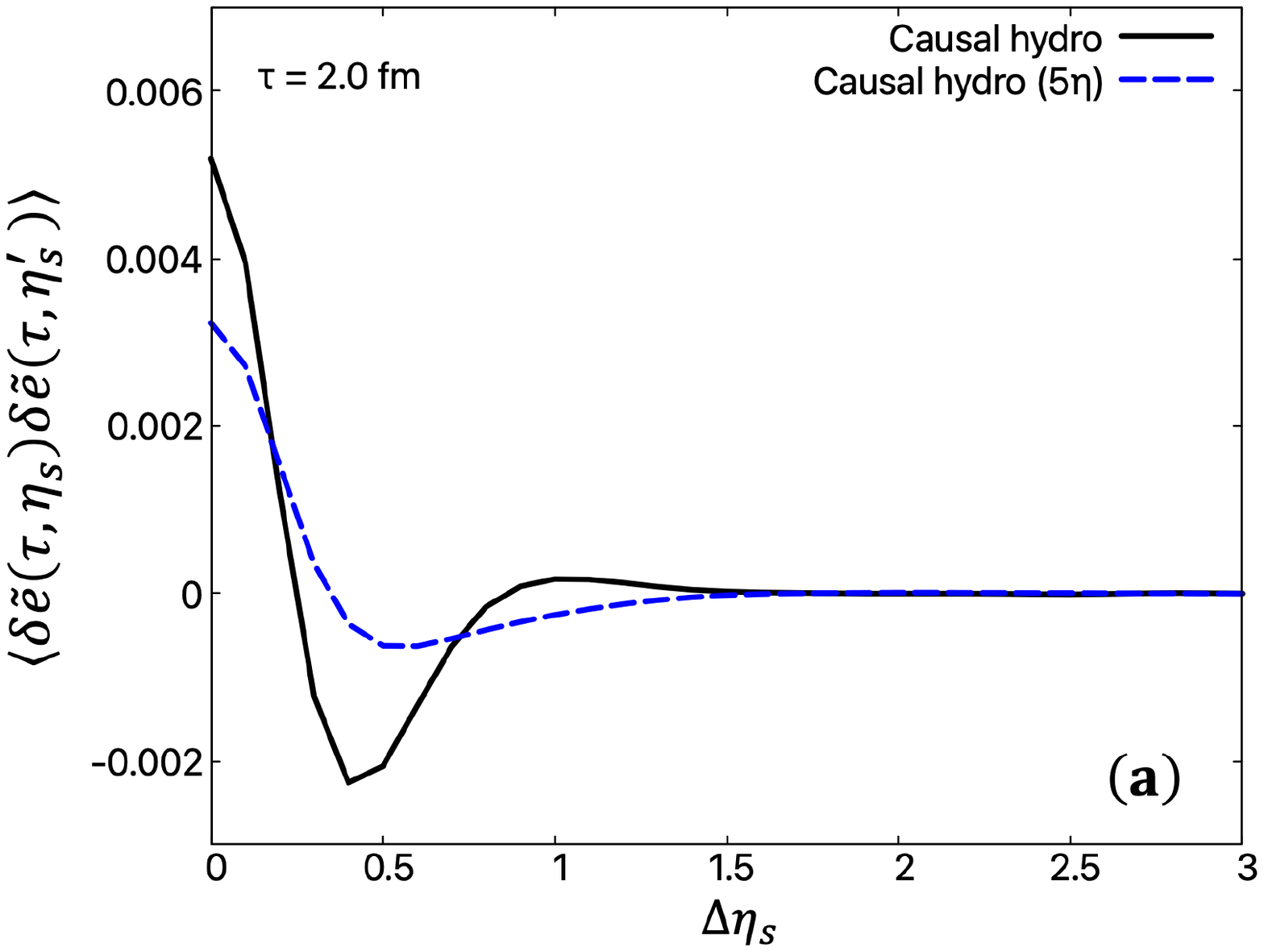}
    \end{minipage}
    \begin{minipage}[b]{1.0\linewidth}
    \centering
    \includegraphics[clip,width=\linewidth]{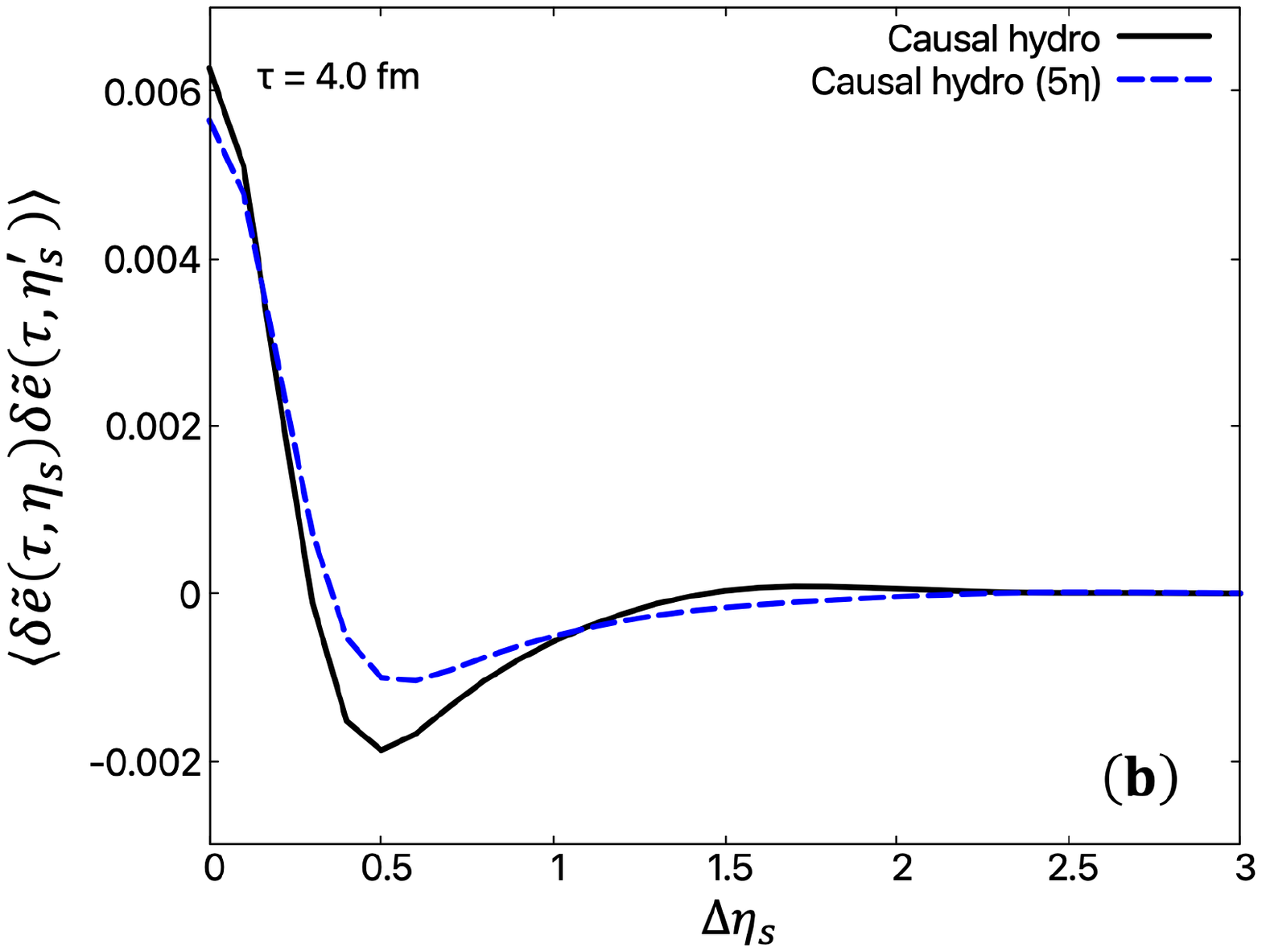}
    \end{minipage}
    \caption{Normalized correlations of energy density fluctuations as functions of $\Delta \eta_{s} \equiv |\eta_{s} - \eta_{s}^{\prime}|$.
    We make a comparison of results from causal hydro with conformal EoS between default settings (black solid line) and 5 times larger $\eta$ (blue dashed line) at (a) $\tau = 2.0 \;\text{fm}$ and (b) $\tau = 4.0 \;\text{fm}$.}
    \label{fig:shear1}
\end{figure}

Next, we investigate correlations of fluctuations and their dependence on various settings.
Figure \ref{fig:order1} shows the relaxation time dependence of two-point correlation functions of normalized energy density at (a) $\tau=1.5$ fm and (b) $\tau = 4.0$ fm. 
The two-point correlation function is defined as
\begin{equation}
\left< \delta\tilde{e}(\tau,\eta_{s}) \delta\tilde{e}(\tau,\eta_{s}^{\prime}) \right> \equiv \frac{\left< \delta e(\tau,\eta_{s}) \delta e(\tau,\eta_{s}^{\prime}) \right>}{e_{0}^{2}(\tau)},
\end{equation}
where angular bracket $\left< \cdots \right>$ means both event average and $\eta_{s}$ average.\footnote{The boost invariant property of background enables us to take the average with respect to space-time rapidity, $\eta_{s}$.
Throughout the present paper, all correlations are averaged over 10,000 events.} 
Regarding the structure of correlations itself, correlations grow up around the origin $\Delta\eta_{s} \approx 0$ and a dip appears at  $\Delta\eta_{s} \approx 0.5$.
This is plausible from a viewpoint of the conservation law: the energy density at some point has a negative correlation with that in its vicinity due to energy conservation.
The acausal scenario can be demonstrated by taking the $\tau_{\pi} \rightarrow 0$ limit in Eqs.~(\ref{0th shear IS final}) and (\ref{1st shear IS final}).
We obviously see a difference among the three different settings of the relaxation time at the early stage, $\tau=1.5$ fm: the correlation with smaller relaxation time of shear pressure tends to propagate faster in the space-time rapidity direction and correlation around the origin becomes stronger.
However, the difference becomes tiny at the late stage, $\tau=4.0$ fm: it takes a longer time to catch up to the final shape of the correlation due to a longer relaxation time.
We also analyze the effect of the shear viscosity on energy density correlations in Fig.~\ref{fig:shear1}. Apparently, fluids with larger shear viscosity tend to behave more slowly and the dip structures are smeared.
\begin{figure}[tbp]
    \begin{minipage}[b]{1.0\linewidth}
    \centering
    \includegraphics[clip,width=\linewidth]{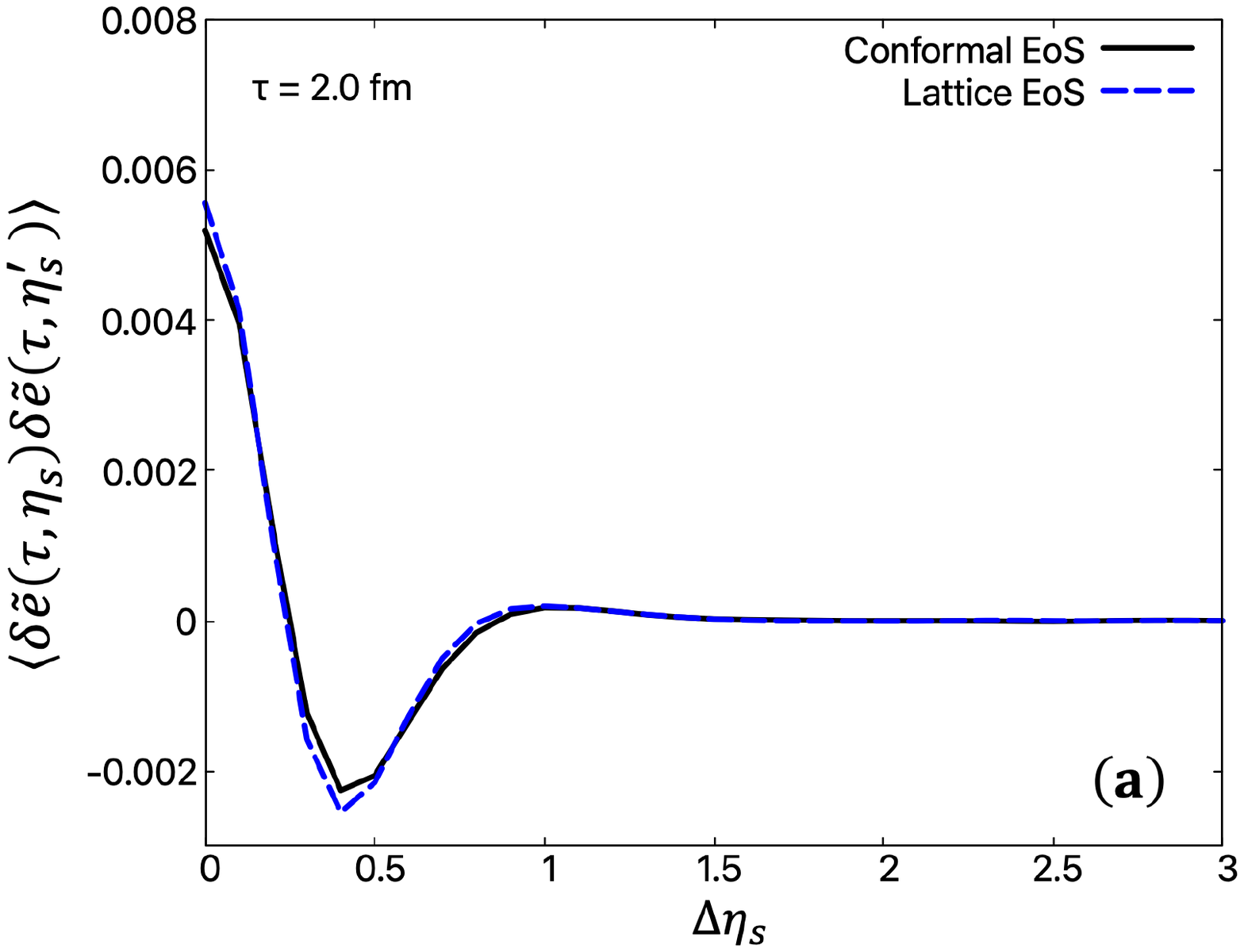}
    \end{minipage}\\
    \begin{minipage}[b]{1.0\linewidth}
    \centering
    \includegraphics[clip,width=\linewidth]{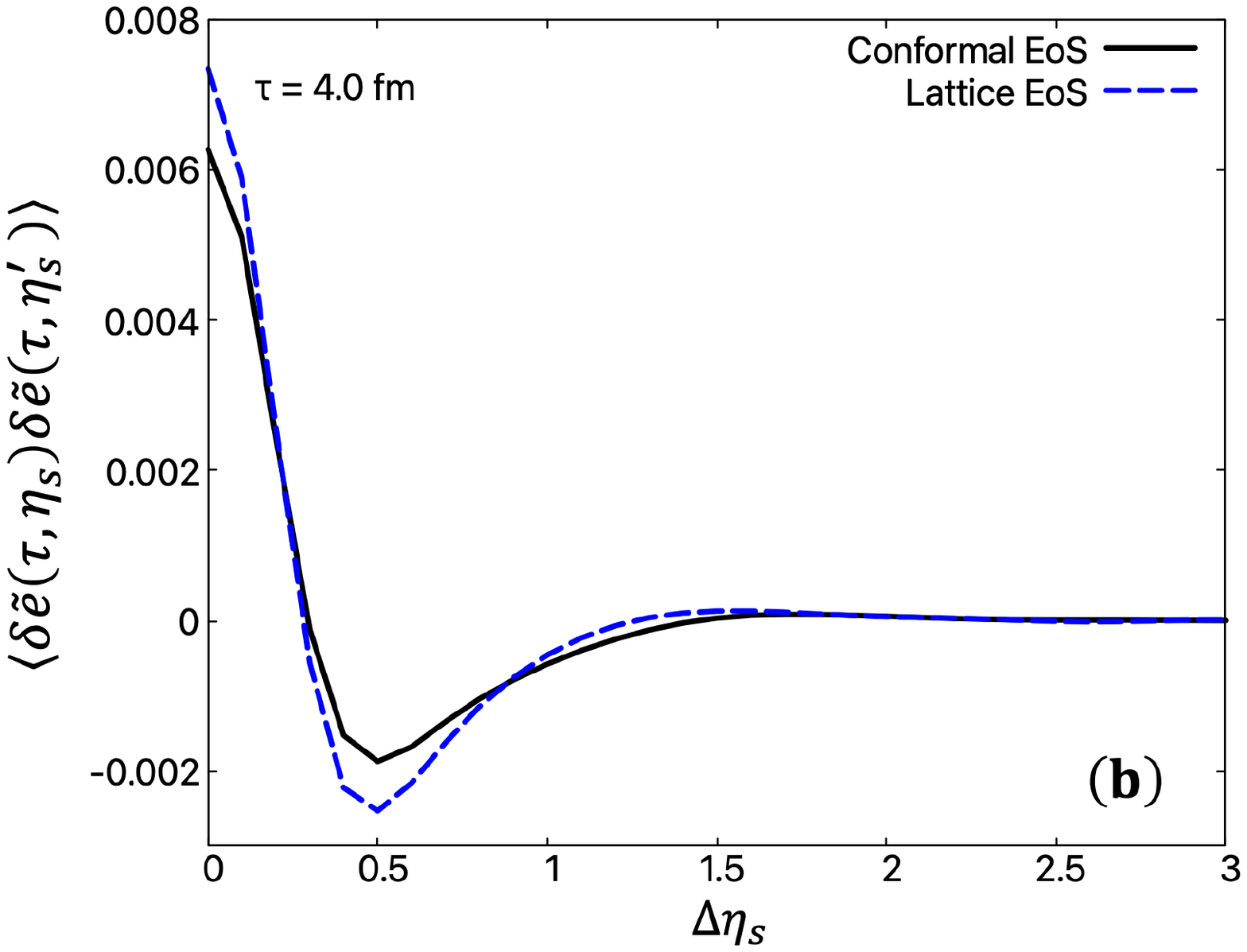}
    \end{minipage}
    \caption{Normalized correlations of energy density fluctuations as functions of $\Delta \eta_{s} \equiv |\eta_{s} - \eta_{s}^{\prime}|$.
    We make a comparison between conformal EoS (default settings, black solid line) and lattice EoS (blue dashed line)
    at  (a) $\tau = 2.0 \;\text{fm}$ and   (b) $\tau = 4.0 \;\text{fm}$. }
    \label{fig:eos1}
\end{figure}

In Fig.~\ref{fig:eos1}, we make a comparison of the correlation between the conformal EoS and the lattice EoS. 
Although the propagation speed of information should be slightly different due to the difference of sound velocity, $c_s^2 = \partial p_0/\partial e_0$, only a small difference can be seen in this comparison.
In general, the sound velocity of lattice EoS is smaller than that of conformal EoS. Hence, information is likely to remain around the origin at which the correlation becomes stronger.
That would be the reason why we observe the difference between these two different EoSs. 
It is noted that the difference in propagation speed can be seen in Fig.~\ref{fig:order1}, which that indicates relaxation time also affects the propagation in the space-time rapidity direction.

\begin{figure}[tbp]
    \begin{minipage}[b]{1.0\linewidth}
    \centering
    \includegraphics[clip,width=\linewidth]{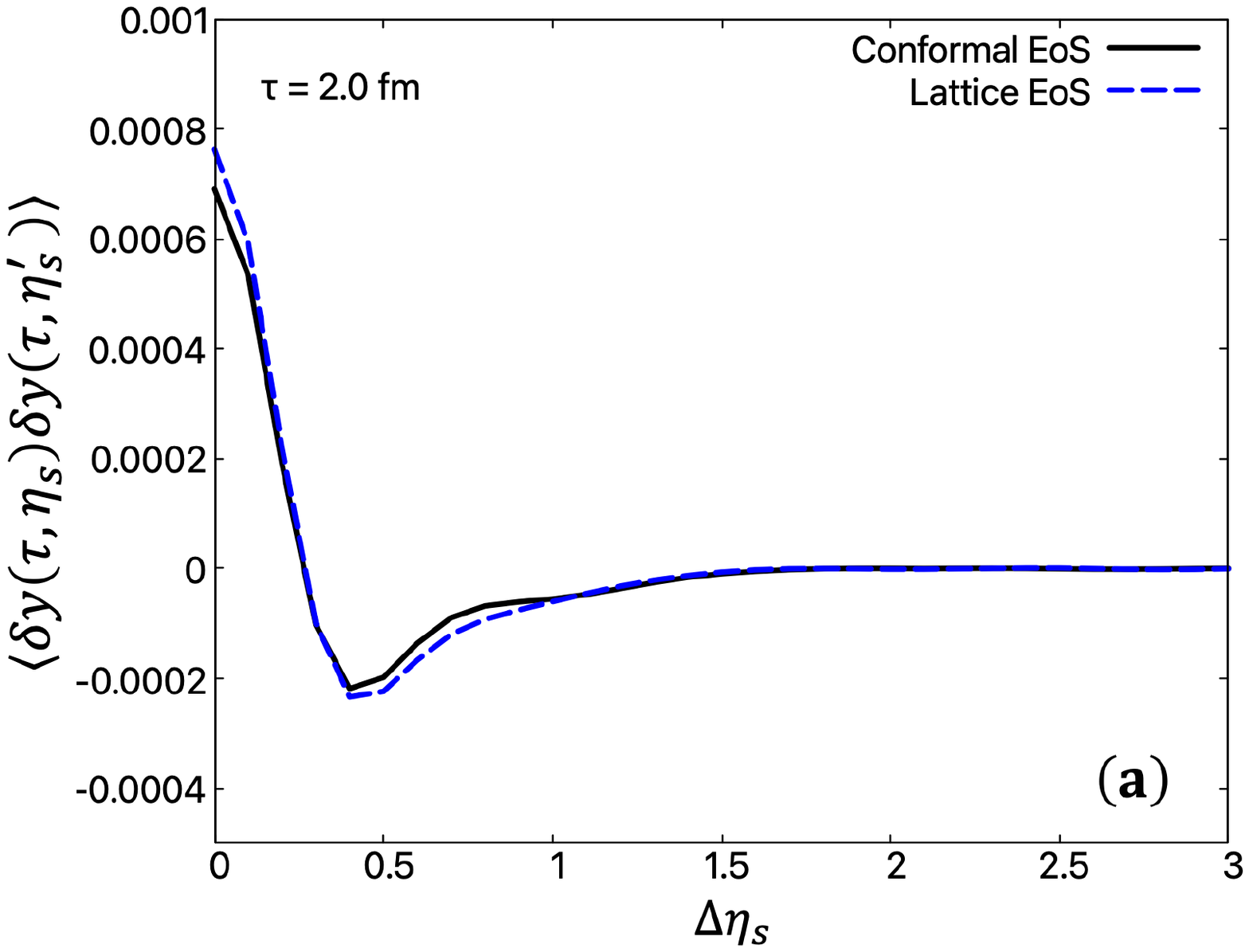}
    \end{minipage}\\
    \begin{minipage}[b]{1.0\linewidth}
    \centering
    \includegraphics[clip,width=\linewidth]{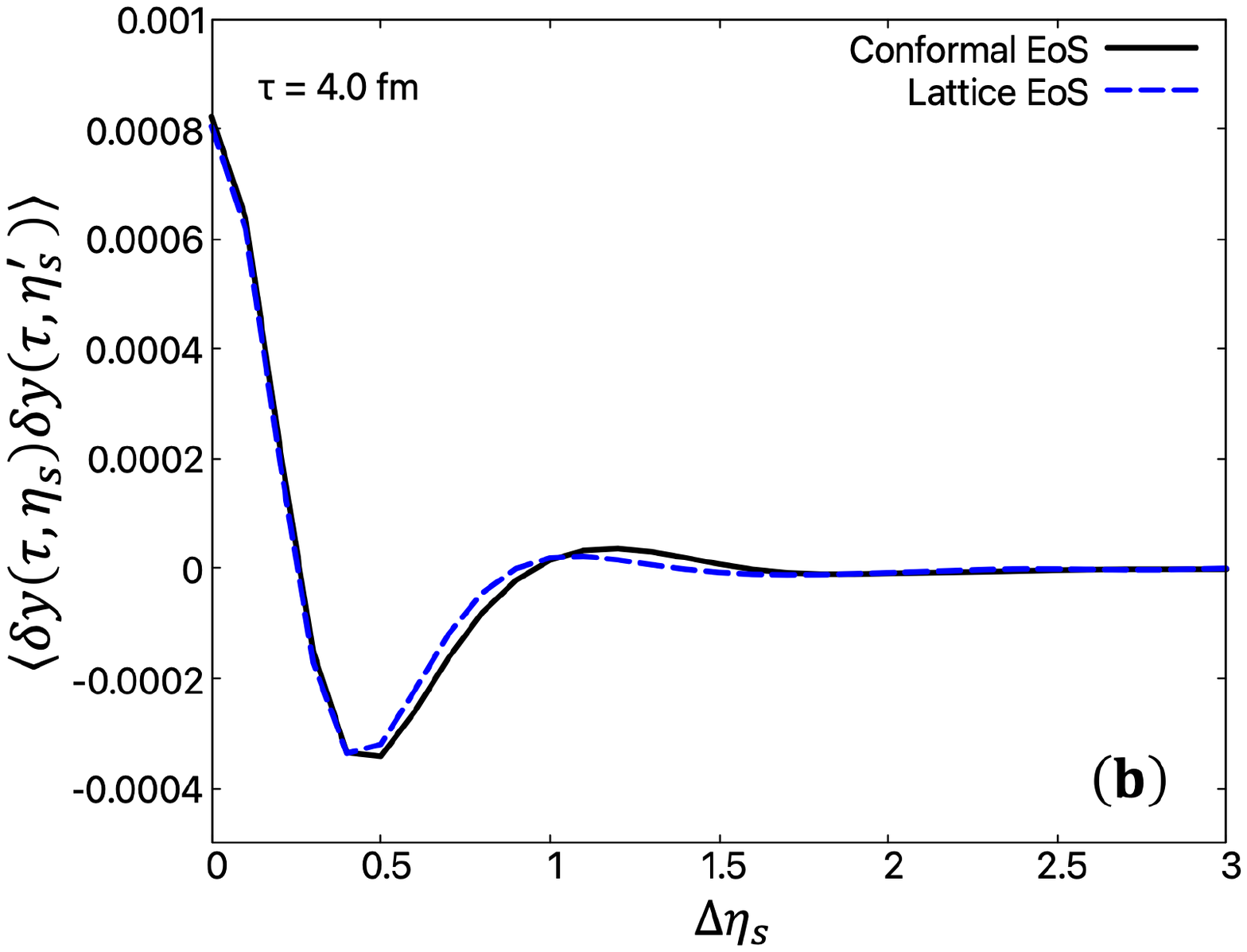}
    \end{minipage}
    \caption{Correlations of flow rapidity fluctuations as functions of $\Delta \eta_{s} \equiv |\eta_{s} - \eta_{s}^{\prime}|$.
    We make a comparison between conformal EoS (default settings, black solid line) and lattice EoS (blue dashed line)
    at  (a) $\tau = 2.0 \;\text{fm}$ and   (b) $\tau = 4.0 \;\text{fm}$. }
    \label{fig:eos1_y}
\end{figure}

\begin{figure}[tbp]
    \begin{minipage}[b]{1.0\linewidth}
    \centering
    \includegraphics[clip,width=\linewidth]{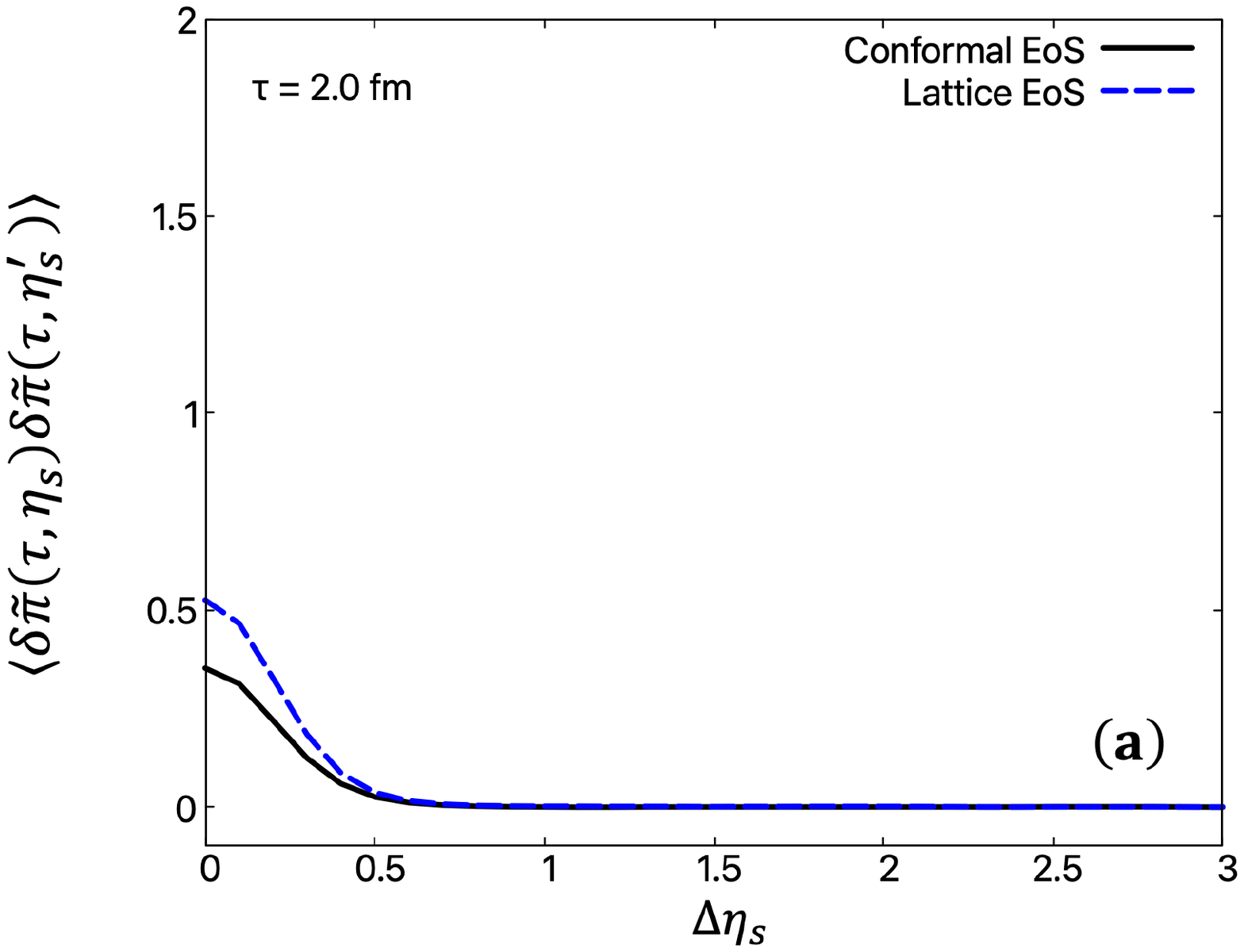}
    \end{minipage}\\
    \begin{minipage}[b]{1.0\linewidth}
    \centering
    \includegraphics[clip,width=\linewidth]{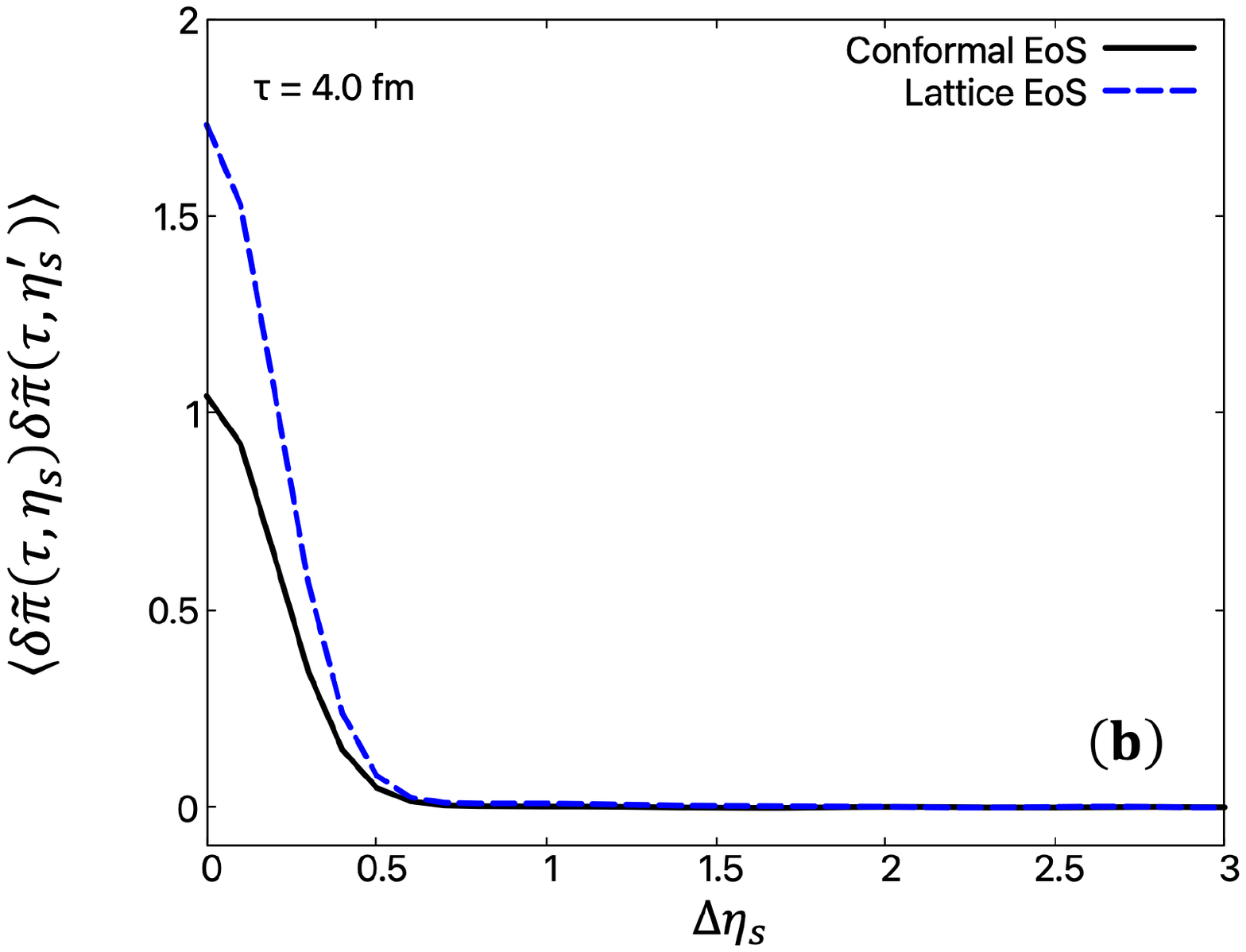}
    \end{minipage}
    \caption{Normalized correlations of shear pressure fluctuations as functions of $\Delta \eta_{s} \equiv |\eta_{s} - \eta_{s}^{\prime}|$.
    We make a comparison between conformal EoS (default settings, black solid line) and lattice EoS (blue dashed line)
    at  (a) $\tau = 2.0 \;\text{fm}$ and   (b) $\tau = 4.0 \;\text{fm}$. }
    \label{fig:eos1_pi}
\end{figure}

We also show correlations of flow rapidity fluctuations in Fig.~\ref{fig:eos1_y} and shear pressure fluctuations in Fig.~\ref{fig:eos1_pi} as comparisons of the effect of model EoS.
The shape of correlations of flow rapidity fluctuations is very similar to that of energy density fluctuations.
It is a reasonable behavior as a variable originating from the conservation law of momentum.
Note that the difference of propagation speed due to the sound velocity difference is relatively visible at $\tau = 4.0 \;\text{fm}$ in Fig.~\ref{fig:eos1_y}: the positions of the second peaks around $\Delta \eta_s \approx 1$ differ.
In Fig.~\ref{fig:eos1_pi}, we do not observe dips and the second peaks in correlations of fluctuations of shear pressure.
The difference in the shape of correlation functions comes from the property of the variable, namely, it is a ``fast variable," which is one of the diffusive quantities and has nothing to do with the conservation law unlike the ``slow variables'' such as energy density and flow rapidity. 
The correlations around the origin monotonically grow up with time evolution and the magnitude is larger in lattice EoS through the whole time evolution.
It can be understood from the viewpoint of the FDR. In fact, the shape of shear pressure correlations directly reflects the FDR since the noise term is introduced in the equation of shear pressure fluctuations (\ref{1st shear IS final}).
From the FDR (\ref{eq:FDR for shear Bj origin}), the magnitude of the noises is larger at higher temperature.
At a fixed proper time, the background temperature under the expansion with lattice EoS is larger than that with conformal EoS since the sound velocity of the former is in general smaller than that of the latter.
Thus, the magnitude of the noises is larger in lattice EoS than in conformal EoS.
This behavior has already been discussed [see Eq.~(\ref{eq:time dependence of noise ratio})] and is also seen in Fig.~\ref{fig:ratio}(c) in Sec.~\ref{sec:validity}.

To summarize these results, shear viscosity and relaxation time work to slow down the behaviors of correlations and also suppress the correlations around the origin and propagation in $\eta_{s}$ space.
Furthermore, the model EoS also affects the behaviors of correlations, which is understood from its sound velocity difference.


\subsection{Two-particle correlation functions}
\label{sec:two-particle correlation}
We have analyzed so far the space-time evolution of thermodynamic variables and their correlations.
We cannot, however, observe them directly in relativistic heavy ion collision experiments.
Therefore it is indispensable to connect these results with experimental observables.
To realize this, we calculate momentum distributions of hadrons using the Cooper-Frye formula \cite{Cooper:1974mv},
\begin{equation}\label{eq:C-F}
E\frac{dN}{d^{3}p} = \frac{d}{(2\pi)^{3}} \int_{\Sigma(x)} p^{\mu}d\sigma_{\mu}(x) \;f \big(p^{\nu}u_{\nu}(x),T(x) \big),
\end{equation}
where $d$, $p^{\mu}=(E, \bm{p})$, $d\sigma_{\mu}(x)$, $\Sigma(x)$, and $f(E, T)$ are degeneracy of a hadron under consideration, four-momentum, normal vector of hypersurface element, particlization hypersurface, and one-particle distribution function, respectively. 
Since we consider finite viscosity, viscous correction to the distribution function should be taken into account \cite{Teaney:2003kp, Monnai:2009ad}.
Thus, we divide the one-particle distribution function into an equilibrium (ideal) part and a near-equilibrium (viscous) part:
\begin{equation}
f = f_{\rm{ideal}} + f_{\rm{vis}}.
\end{equation}
For simplicity, we assume Boltzmann distribution for an equilibrium part $f_{\rm{ideal}}$ and, within our framework, it is written as
\begin{equation}\label{eq:Boltzmann}
f_{\rm{ideal}}(p_T, Y; \tau,\eta_{s}) = \text{exp} \left[-\frac{m_{T}\cosh\left( Y-\eta_{s}-\delta y(\tau,\eta_{s}) \right)}{T_{0}(\tau) + \delta T(\tau,\eta_{s})}\right],
\end{equation}
where $Y\equiv\tanh^{-1}(p_{z}/E)$, $m_{T}\equiv\sqrt{p_{T}^{2}+m^{2}}$, $T_{0}$, and $\delta T$ are rapidity of particles, transverse mass, temperature of background, and fluctuation of temperature, respectively.\footnote{Temperature of background, $T_{0}$,  and fluctuation of temperature, $\delta T$, are calculated from $e_{0}$ and $\delta e$  for a given EoS.}
On the other hand, a near-equilibrium part \cite{Monnai:2009ad} is written as
\begin{align}
&f_{\rm{vis}}(p_T, Y; \tau,\eta_{s}) \nonumber \\
&= \frac{1}{2sT^{3}}f_{\rm{ideal}}p^{\mu}p^{\nu}\pi_{\mu\nu}\nonumber\\
&= \frac{1}{2(s_{0}+\delta s)(T_{0}+\delta T)^{3}}f_{\rm{ideal}}\nonumber\\
&\quad \times \left[ \frac{1}{2}p_{T}^{2} - m_{T}^{2} \sinh^{2}(Y - \eta_{s} - \delta y) \right](\pi_{0}+\delta \pi),
\end{align}
where $s$ is entropy density and $s_{0}$ and $\delta s$ are its background and fluctuation, respectively.
Furthermore, we consider perturbative expansion of $f_{\rm{ideal}}$ and $f_{\rm{vis}}$ with respect to the fluctuations $\delta y$, $\delta T$, $\delta s$, and $\delta \pi$ up to the first order following the same prescription as derivation of EoMs of fluctuations in Sec.~\ref{sec:model}.
The resultant distributions are,
\begin{align}
f_{\rm{ideal}} &\approx e^{-\frac{m_{T}\cosh\left( Y-\eta_{s} \right)}{T_{0}}}\nonumber\\
&\quad + e^{-\frac{m_{T}\cosh\left( Y-\eta_{s} \right)}{T_{0}}} \left[\frac{m_{T}\sinh(Y-\eta_{s})}{T_{0}}\delta y \right. \nonumber \\
&\quad \left.+ \frac{m_{T}\cosh(Y-\eta_{s})}{T_{0}^{2}}\delta T  \right] \nonumber\\
&\equiv f_{\rm{ideal},0} + \delta f_{\rm{ideal}},\label{eq:fideal}
\end{align}
where $f_{\rm{ideal},0}$ and $\delta f_{\rm{ideal}}$ are the zeroth- and the first-order perturbative terms of an equilibrium part, respectively.
The perturbative expansion of near-equilibrium parts leads to a much more complicated form:
\begin{align}
f_{\rm{vis}} &\approx e^{-\frac{m_{T}\cosh\left( Y-\eta_{s} \right)}{T_{0}}}
 \frac{1}{2s_{0}T_{0}^{3}} \left[ \frac{1}{2}p_{T}^{2} - m_{T}^{2}\sinh^{2}(Y-\eta_{s}) \right]\pi_{0}\nonumber\\
&\quad+e^{-\frac{m_{T}\cosh\left( Y-\eta_{s} \right)}{T_{0}}} \frac{1}{2s_{0}T_{0}^{3}}\nonumber\\
&\quad\times \left\{\left[ \frac{m_{T}\sinh(Y-\eta_{s})}{T_{0}}\delta y + \frac{m_{T}\cosh(Y-\eta_{s})}{T_{0}^{2}}\delta T \right] \right. \nonumber\\
&\quad\times \left[ \frac{1}{2}p_{T}^{2} - m_{T}^{2}\sinh^{2}(Y-\eta_{s}) \right]\pi_{0}\nonumber\\ 
&\quad- \frac{1}{T_{0}}\left(\frac{1}{c_{s}^{2}} + 3\right)\left[ \frac{1}{2}p_{T}^{2} - m_{T}^{2}\sinh^{2}(Y-\eta_{s}) \right]\pi_{0} \delta T\nonumber\\
&\quad+ 2m_{T}^{2}\sinh(Y-\eta_{s})\cosh(Y-\eta_{s})\pi_{0}\delta y\nonumber\\
&\quad+  \left.\left[ \frac{1}{2}p_{T}^{2} - m_{T}^{2}\sinh^{2}(Y-\eta_{s}) \right] \delta \pi \right\}\nonumber\\
&\equiv f_{\rm{vis},0} + \delta f_{\rm{vis}},\label{eq:fvis}
\end{align}
where $f_{\rm{vis},0}$ and $\delta f_{\rm{vis}}$ are the zeroth- and the first-order perturbative terms of a near-equilibrium part, respectively.
To summarize, we have
\begin{equation}\label{eq:expansion}
f = f_{\rm{ideal}} + f_{\rm{vis}} = f_{\rm{ideal},0} + \delta f_{\rm{ideal}} + f_{\rm{vis},0} + \delta f_{\rm{vis}}
\end{equation}
as a one-particle distribution function to calculate momentum distributions via the Cooper-Frye formula (\ref{eq:C-F}).

Rapidity distribution of hadrons is obtained by using the relations $d^{3}p/E = d^{2}p_{T}dY = p_{T} dp_{T} dY d\phi= m_{T}dm_{T} dY d\phi$, and by integrating over azimuthal angle:
\begin{align}
\frac{dN}{dY}  =& \frac{d}{(2\pi)^{2}} \tau A\int_{-\infty}^{\infty}d\eta_{s} \int_{m}^{\infty}dm_{T}m_{T}^{2}\cosh(Y-\eta_{s}) \; f, \label{eq:rapidity dist}
\end{align}
where $A$ and $m$ are transverse area and mass of a hadron under consideration, respectively.
In the above calculation, we assumed isochronous freeze-out at $\tau$ and also used the relation $p^{\mu}d\sigma_{\mu} = \tau A d\eta_{s} m_{T} \cosh(Y-\eta_{s})$ by neglecting possible small changes of freeze-out hypersurface due to fluctuations.
By taking an average of Eq.~(\ref{eq:rapidity dist}) over events and integrating over $m_{T}$, the first-order perturbative parts in the one-particle distribution function vanish.
We finally obtain
\begin{widetext}
\begin{align}\label{eq:one particle correlation}
\left< \frac{dN}{dY} \right> &=  \frac{d}{(2\pi)^{2}} \tau  A\int_{Y - \Delta\eta_{s}^{\prime}}^{Y + \Delta\eta_{s}^{\prime}}d\eta_{s} \cosh{(Y-\eta_{s})}
e^{-\frac{m\cosh\left( Y-\eta_{s} \right)}{T_{0}}} \Biggl\{ \left(1-\frac{1}{2s_{0}T_{0}^{3}}\frac{1}{2}m^{2}\pi_{0} \right)\nonumber\\
&\quad\times \Biggl[ \frac{T_{0}m^{2}}{\cosh(Y-\eta_{s})} + \frac{2T_{0}^{2}m}{\cosh^{2}(Y-\eta_{s})} + \frac{2T_{0}^{3}}{\cosh^{3}(Y-\eta_{s})} \Biggr]
+ \frac{1}{2s_{0}T_{0}^{3}}\left( \frac{1}{2}-\sinh^{2}(Y-\eta_{s}) \right)\pi_{0}\nonumber\\
&\quad\times \Biggl[ \frac{T_{0}m^{4}}{\cosh(Y-\eta_{s})} + \frac{4T_{0}^{2}m^{3}}{\cosh^{2}(Y-\eta_{s})} + \frac{12T_{0}^{3}m^{2}}{\cosh^{3}(Y-\eta_{s})}
+ \frac{24T_{0}^{4}m}{\cosh^{4}(Y-\eta_{s})} + \frac{24T_{0}^{5}}{\cosh^{5}(Y-\eta_{s})} \Biggr] \Biggr\}.
\end{align}
\end{widetext}
Practically, we replaced the integral range of $\eta_{s}$, $[-\infty, \infty]$, with $[Y - \Delta \eta_{s}^{\prime}, Y + \Delta \eta_{s}^{\prime}]$ since, for a given $Y$, a contribution from a fluid element being far away from the one at $\eta_{s}=Y$ can be negligible.
We set $\Delta \eta_{s}^{\prime} = 3$ in our calculations and confirmed that this range is sufficient enough to converge the integral.
The first-order terms, $\delta f_{\rm{ideal}}$ and $\delta f_{\rm{vis}}$, do not contribute to the final one-particle momentum distribution due to $\left< \delta f \right> = 0$.
It is noted that, in the derivation of Eq.~(\ref{eq:one particle correlation}), an integral with respect to transverse mass is analytically solved using the incomplete gamma function (see also Appendix \ref{sec:B}).

Two-particle correlation functions in the rapidity direction, $\left< \frac{dN}{dY_{1}}\frac{dN}{dY_{2}} \right>$, are also obtained from the Cooper-Frye formula (\ref{eq:C-F}) after straightforward but lengthy calculations.\footnote{Here we neglect possible quantum correlations between two identical particles.}
Here subscripts 1 and 2 are labels of particles 1 and 2, respectively.
See also Appendix \ref{sec:B} for details of two-particle correlation functions.
Finally, we obtain the normalized two-particle correlations $\left< \frac{dN}{dY_{1}}\frac{dN}{dY_{2}} \right> / \left< \frac{dN}{dY_{1}} \right> \left< \frac{dN}{dY_{2}} \right>$ as a function of rapidity gap $\Delta Y \equiv |Y_{1}-Y_{2}|$.
In what follows, the default model of the EoS is the lattice EoS although we employ the conformal EoS for the sake of comparison.\footnote{In the conformal EoS, the constituents of the fluids should be massless particles.
Nevertheless, we employ it and calculate spectra for massive particles to capture the difference in hydrodynamic evolution between conformal EoS and lattice EoS.}

\begin{figure}[tbp]
    \centering
    \includegraphics[clip,width=\linewidth]{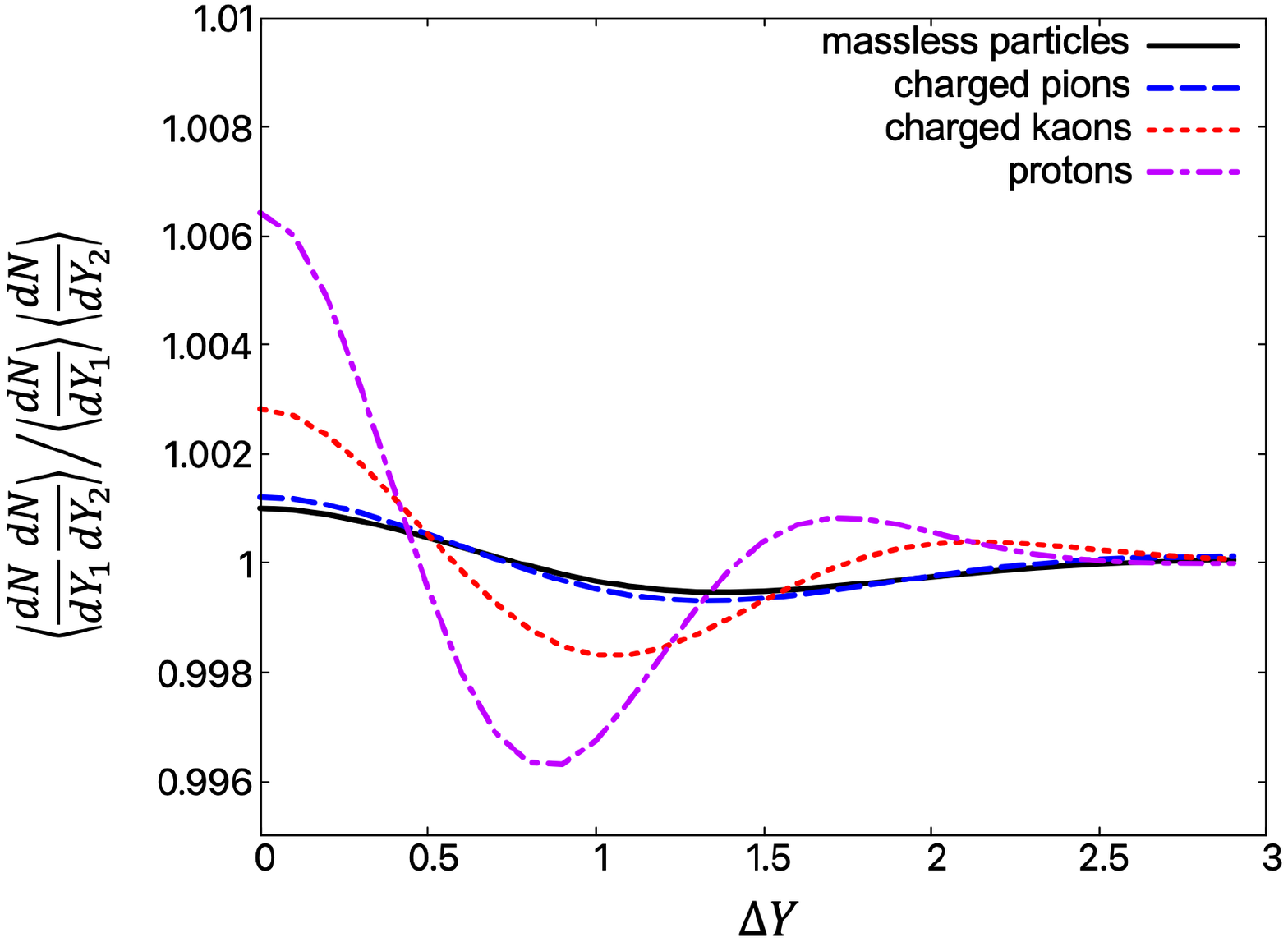}
    \caption{Normalized two-particle correlation functions as functions of  $\Delta Y \equiv |Y_{1} - Y_{2}|$ for massless particles  (black solid line), charged pions (blue dashed line), charged kaons (red dotted line), and protons  (magenta dash-dotted line) at freeze-out time $\tau = 10.0 \;\text{fm}$ and the corresponding background temperature $T_{0} \simeq 0.168 \;\text{GeV}$. Lattice EoS is employed for hydrodynamic evolution.
    }
    \label{fig:particle_correlation_mass}
\end{figure}

Figure \ref{fig:particle_correlation_mass} shows the normalized two-particle correlation functions as functions of rapidity gap $\Delta Y$ for massless particles ($m=0$ GeV), charged pions ($m=0.139$ GeV), charged kaons ($m=0.494$ GeV), and protons ($m=0.938$ GeV) without any contributions from resonance decays.
Lattice EoS and default setting for transport coefficients are employed to describe the hydrodynamic evolution in this analysis.
We assume that all hadrons freeze out at $\tau = 10.0 \;\text{fm}$ which corresponds to freeze-out temperature $T\simeq 0.168$ GeV in the lattice EoS and transport coefficients employed in this study.
Since the two-particle correlation functions have a value rather than unity when two-point correlation functions such as $\langle \delta T(\eta_{s1})\delta T(\eta_{s2})\rangle$ and $\langle \delta y(\eta_{s1})\delta y(\eta_{s2})\rangle$ are finite, it is obvious that the information at freeze-out is inherited by two-particle correlation functions.
Moreover, the pattern of the correlations is more clearly seen for heavier hadrons.
It indicates that the heavier hadrons are good probes of correlations and that they can be used to extract the properties of the expanding media.
As one sees, positions and depths (heights) of dips (bumps) of correlations depend on the mass of hadrons.
One can interpret this behavior from the viewpoint of thermal fluctuations.
The ratio $m/T_{0}$ appearing in Eq.~(\ref{eq:one particle correlation}) is a good measure of thermal fluctuations in a fluid element and characterizes the shape of correlations: the momentum rapidity, $Y$, of heavier hadrons tends to reflect the flow rapidity, $y_f$, at freeze-out, namely $Y \approx y_f$, while the pattern that correlations of thermodynamic variables possess in the space-time rapidity is blurred by thermal fluctuations in the correlations of lighter particles in momentum rapidity space.
\begin{figure}[tbp]
    \centering
    \includegraphics[clip,width=\linewidth]{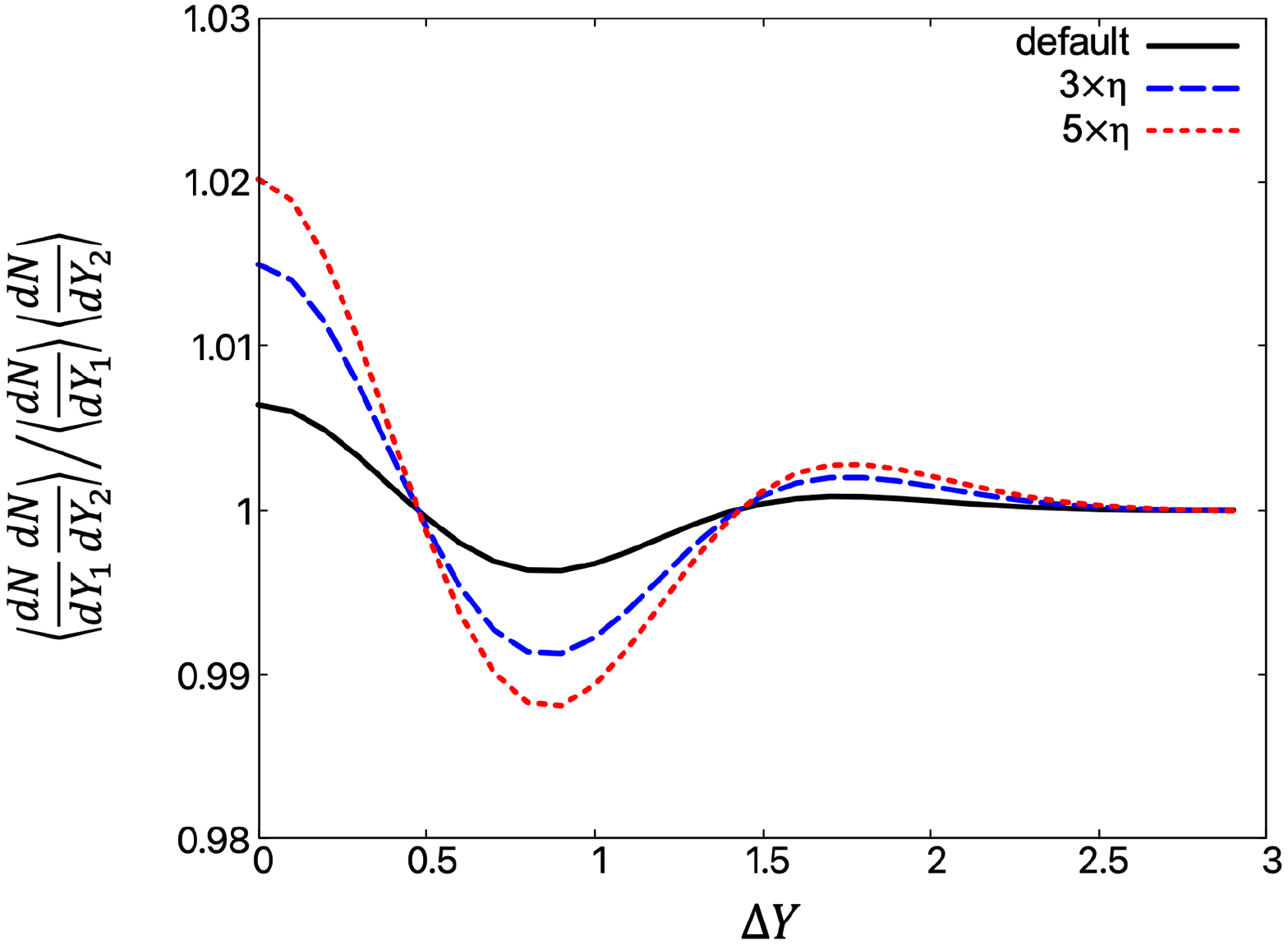}
    \caption{Normalized two-particle correlation functions of protons as functions of $\Delta Y \equiv |Y_{1} - Y_{2}|$ for default set of shear viscosity $\eta$ (black solid line), 3 times larger $\eta$ (blue dashed line), and 5 times larger $\eta$ (red dotted line) at freeze-out time $\tau = 10.0 \;\text{fm}$. Lattice EoS and the default relaxation time are employed for hydrodynamic evolution.
    }
    \label{fig:particle_correlation_shear}
\end{figure}
\begin{figure}[tbp]
    \centering
    \includegraphics[clip,width=\linewidth]{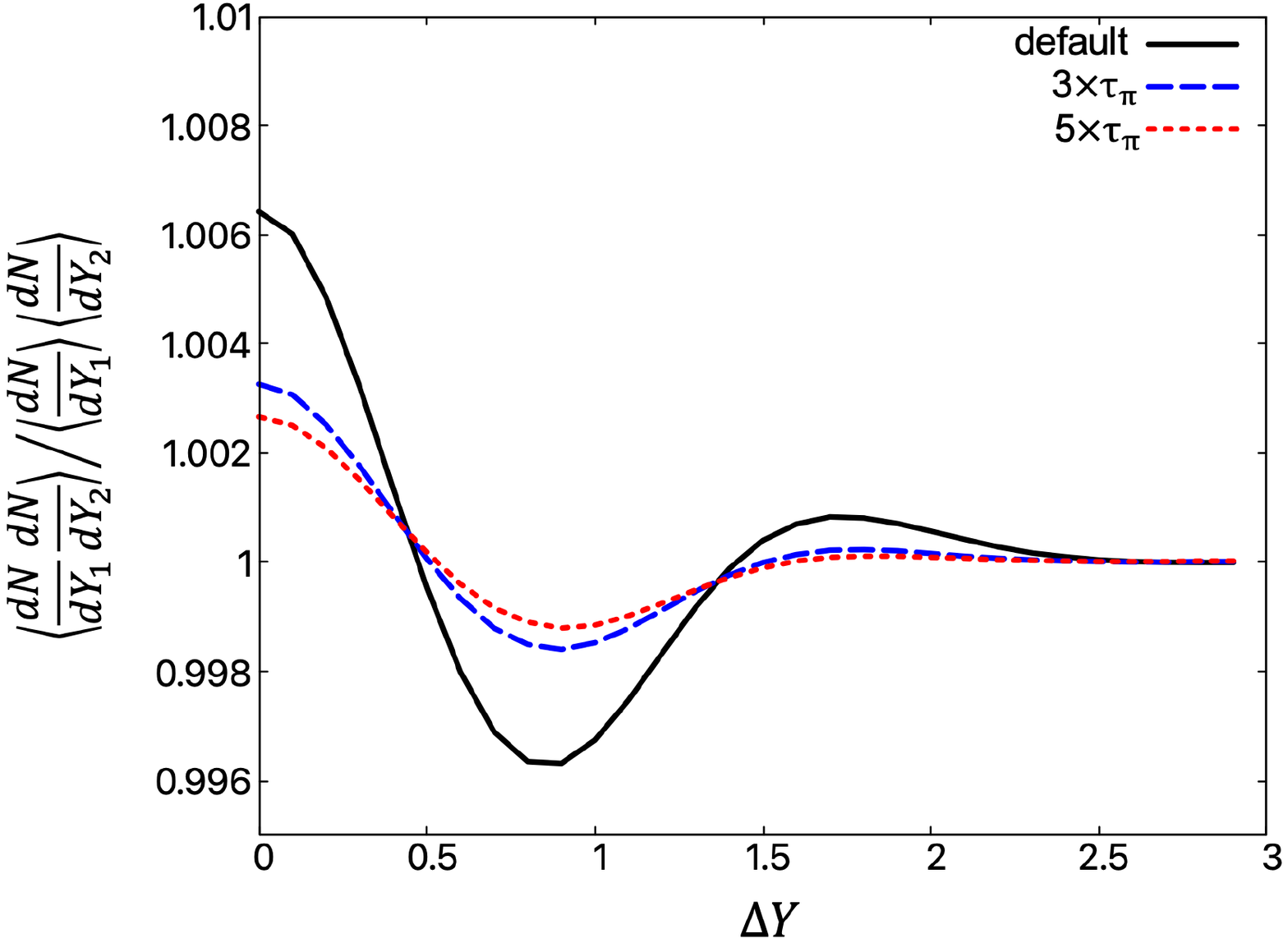}
    \caption{Normalized two-particle correlation functions of protons as functions of $\Delta Y \equiv |Y_{1} - Y_{2}|$ for default set of relaxation time $\tau_{\pi}$ (black solid line), 3 times larger $\tau_{\pi}$ (blue dashed line), and 5 times larger $\tau_{\pi}$ (red dotted line) at freeze-out time $\tau = 10.0 \;\text{fm}$. Lattice EoS and the default shear viscosity are employed for hydrodynamic evolution.
    }
    \label{fig:particle_correlation_relax}
\end{figure}
\begin{figure}[tbp]
    \centering
    \includegraphics[clip,width=\linewidth]{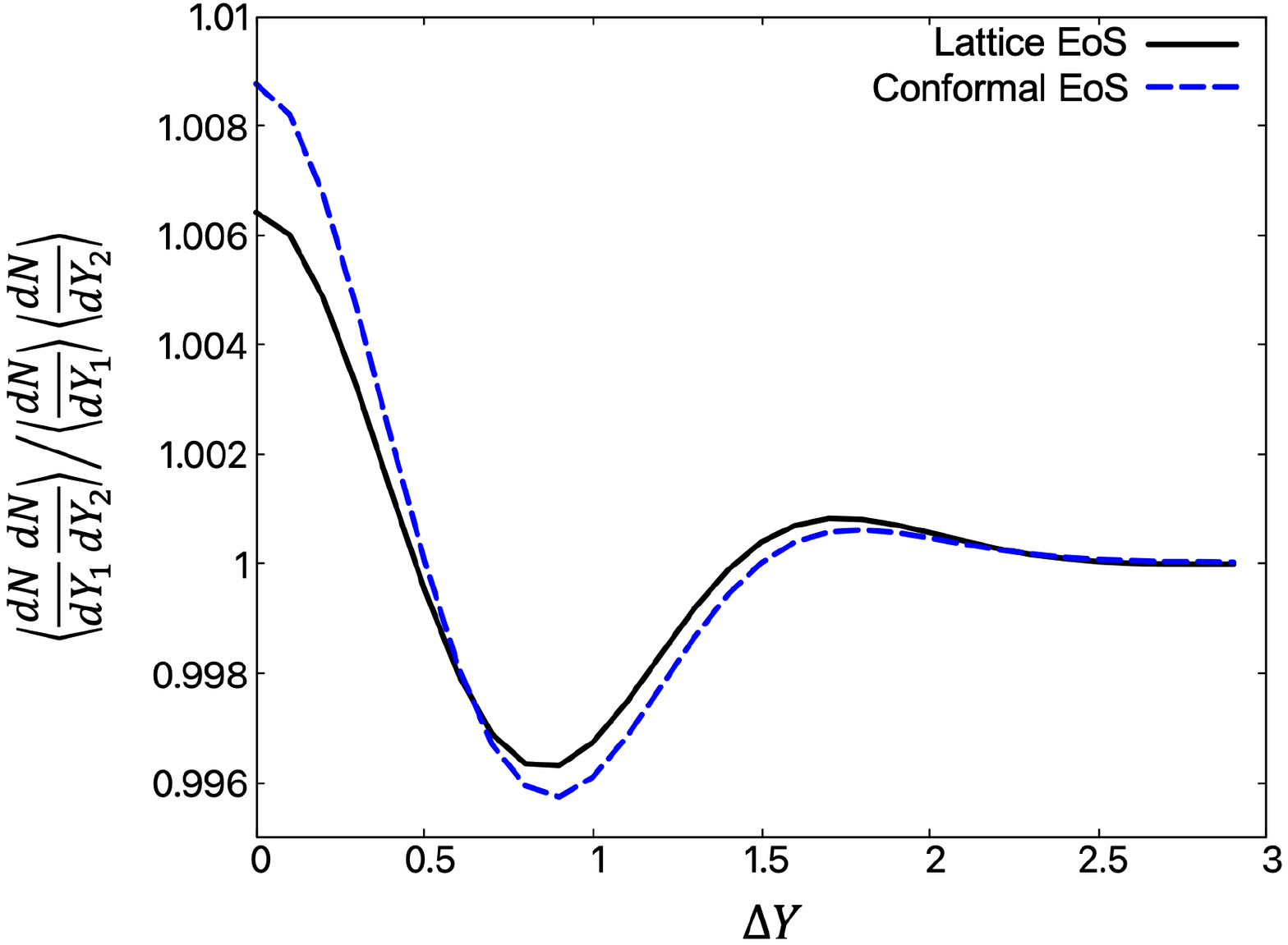}
    \caption{Normalized two-particle correlations of protons as functions of $\Delta Y \equiv |Y_{1} - Y_{2}|$.
    We make a comparison of the results between lattice EoS (black solid line) and conformal EoS (blue dashed line) at $T_{0} \simeq 0.168 \;\text{GeV}$. 
    Regarding the difference in freeze-out time, see the text.}
    \label{fig:particle_correlation_EoS}
\end{figure}

In Figs.~\ref{fig:particle_correlation_shear} and \ref{fig:particle_correlation_relax}, we elucidate how the two-particle correlation functions of protons depend on the transport coefficients under the space-time evolution with lattice EoS.
The correlation pattern becomes more visible by increasing shear viscosity as shown in Fig.~\ref{fig:particle_correlation_shear}.
On the other hand, the magnitude decreases with increasing relaxation time as shown in Fig.~\ref{fig:particle_correlation_relax}.
The magnitude of two-particle correlation functions turns out to be highly sensitive to the transport properties of the media although positions of dips and bumps of the two-particle correlations do not depend on the choice of shear viscosity and relaxation time within a certain range. 
This clearly exhibits that the two-particle correlation functions of relatively heavy hadrons open up a new window to study the transport properties of the media created in relativistic heavy ion collisions.

In Fig.~ \ref{fig:particle_correlation_EoS}, we make a comparison of the proton two-particle correlation functions between conformal EoS and lattice EoS.
When we analyze the correlation functions of energy density fluctuations, the magnitude of lattice EoS is larger than that of conformal EoS as shown in Fig.~\ref{fig:eos1}.
However, this behavior is opposite in the two-particle correlation functions of protons: the magnitude of two-particle correlation functions of protons with conformal EoS is larger than that of lattice EoS.
This is due to the difference in freeze-out time.
In fact, fluids with conformal EoS reach $T_{0} \simeq 0.168 \;\text{GeV}$ faster than the ones with lattice EoS.
Thus, freeze-out time depends on the choice of the model EoS: $\tau = 4.3$ fm for conformal EoS with the degree of freedom $d=47.5$, and $\tau = 10.0$ fm for lattice EoS.   
Thus two-particle correlation functions with conformal EoS reflect the early time correlations of thermodynamic variables which are in the course of growing rapidly.

\section{Summary}
\label{sec:summary}
In this paper, we developed a framework which deals with causal hydrodynamic fluctuations in a one-dimensional expanding system.
Since the EoMs are derived without any assumption of specific constitutive equations, one employs any kinds of constitutive equations proposed so far for shear pressure $\pi$ and bulk pressure $\Pi$.
In the present work, we employed the simplest second-order constitutive equations derived by Israel and Stewart which satisfy the causality.
We solved a system of stochastic partial differential equations and analyzed the dynamics of (1+1)-dimensional causal hydrodynamic fluctuations on an event-by-event basis.
Through the description of space-time evolution of thermodynamic variables, we found a novel phenomenon, an almost frozen structure of energy density fluctuations and flow rapidity fluctuations.
From the analysis of the structure, it could be possible to extract information of the early stage of hydrodynamic evolution, such as thermal fluctuations. 
We also investigated the two-point correlation functions of energy density fluctuations which are induced by hydrodynamic fluctuations and saw the dependence on various settings of the bulk and the transport properties.
The correlations are sensitive to the settings of EoS and transport coefficients such as relaxation time and viscosity.
To extract properties of the media in comparison with experimental data, it is also important to analyze the effects of correlations of thermodynamic variables on two-particle correlation functions which are observables in experiments.
Utilizing the Cooper-Frye formula, we analytically derived the specific form of two-particle correlation functions including two-point correlation functions of thermodynamic variables, flow rapidity, shear pressure, and cross-terms among them.
We show that the two-point correlations of these hydrodynamic variables are inherited by two-particle correlations in the final state.
We found that correlations are more enhanced for heavier particles.
It indicates that heavier hadrons are good probes to see the consequences of hydrodynamic fluctuations.
We also show that the two-particle correlation functions are sensitive to the properties of the medium.
Specifically, the correlations enhance with increasing shear viscosity or decreasing relaxation time.
Sound velocity also affects the shape of correlations, therefore we have a chance to extract the information of the EoS through analysis of the two-particle correlation functions.
These results provide us with an opportunity for multidimensional analysis of properties of the QGP created in relativistic heavy ion collisions.

In the present paper, we assumed one of the simplest settings, e.g., the simplest one-dimensional expanding system, the simplest EoS, the simplest causal constitutive equations, and we neglect charge current and bulk pressure.
Therefore, there is a lot of room for sophistication and extension.
As one way of major extension, the (3+1)-dimensional expanding analytic solution  \cite{Gubser:2010ze}, instead of the one-dimensional boost invariant solution (\ref{eq:Bj solution}), could be employed by following the same prescription that we explained in this paper.
Regarding the FDRs (\ref{eq:FDR for shear Bj}) and (\ref{eq:FDR for bulk Bj}), it is known that these FDRs need additional correction terms when the background system is nonstatic and constitutive equations have finite relaxation effects \cite{Murase:2019cwc}.
It would be intriguing to see the effects of expansion and finite relaxation on the two-point correlations of hydrodynamic variables.
Including fluctuations of conserved charges \cite{Ling:2013ksb,Kapusta:2017hfi,De:2020yyx} in the present framework might be also interesting.
Through the description of causal propagation and diffusion of charge fluctuations, we may have a chance to extract the diffusion coefficients of conserved charges. Since it is beyond the scope of the present study, we leave it for future work.

The present framework can be utilized to search for the critical point of the QCD phase diagram through the analysis of detailed dynamics of both hydrodynamic and critical fluctuations.
Fluctuations of the order parameter of chiral phase transition in the vicinity of the critical point are ``fast modes,'' therefore such information is likely to be lost due to the diffusive mode in the QGP. 
The fluctuations of the order parameter are, however, taken over by the ``slow modes,'' namely, fluctuations of baryon number density \cite{Fujii:2004jt,Son:2004iv}.
Chasing the dynamics of both the slow modes and the fast mode \cite{Sakai:2023lym}, we can diagnose the QCD phase diagram and have a chance to pin down the location of the critical point, but we leave it for future study.

\section*{Acknowledgement}
We would like to thank C.~Chattopadhyay for giving us some useful references.
We also thank S.~Jeon for giving us some valuable comments and Y.~Kanakubo and K.~Kuroki for fruitful discussions.
The work by T.H. was partly supported by JSPS KAKENHI Grant No.~JP19K21881.

\appendix
\section{Sound propagation}\label{sec:A}
\begin{figure}[htbp]
    \begin{minipage}[b]{1.0\linewidth}
    \centering
    \includegraphics[clip,width=\linewidth]{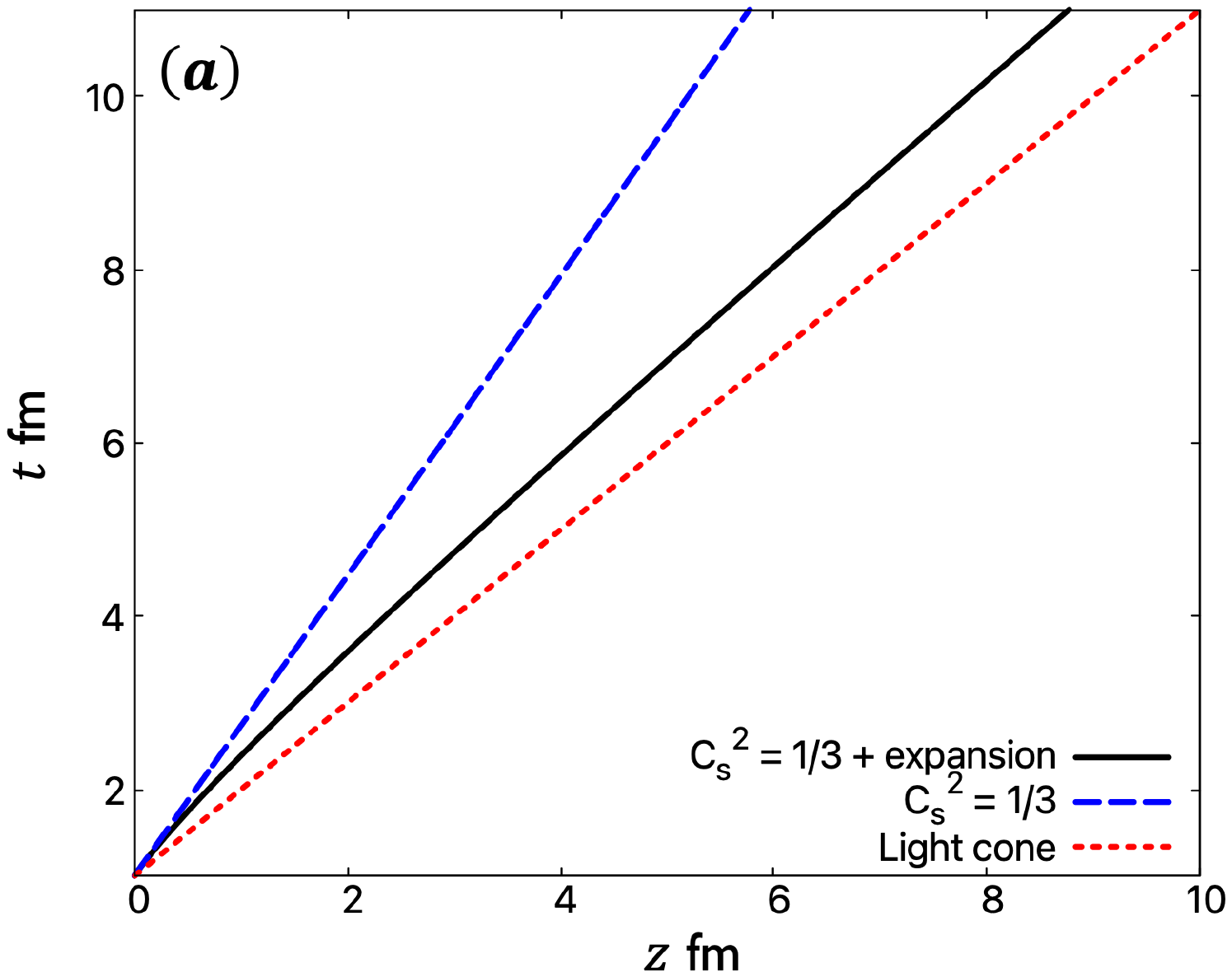}\\
    \end{minipage}\\
    \begin{minipage}[b]{1.0\linewidth}
    \centering
    \includegraphics[clip,width=\linewidth]{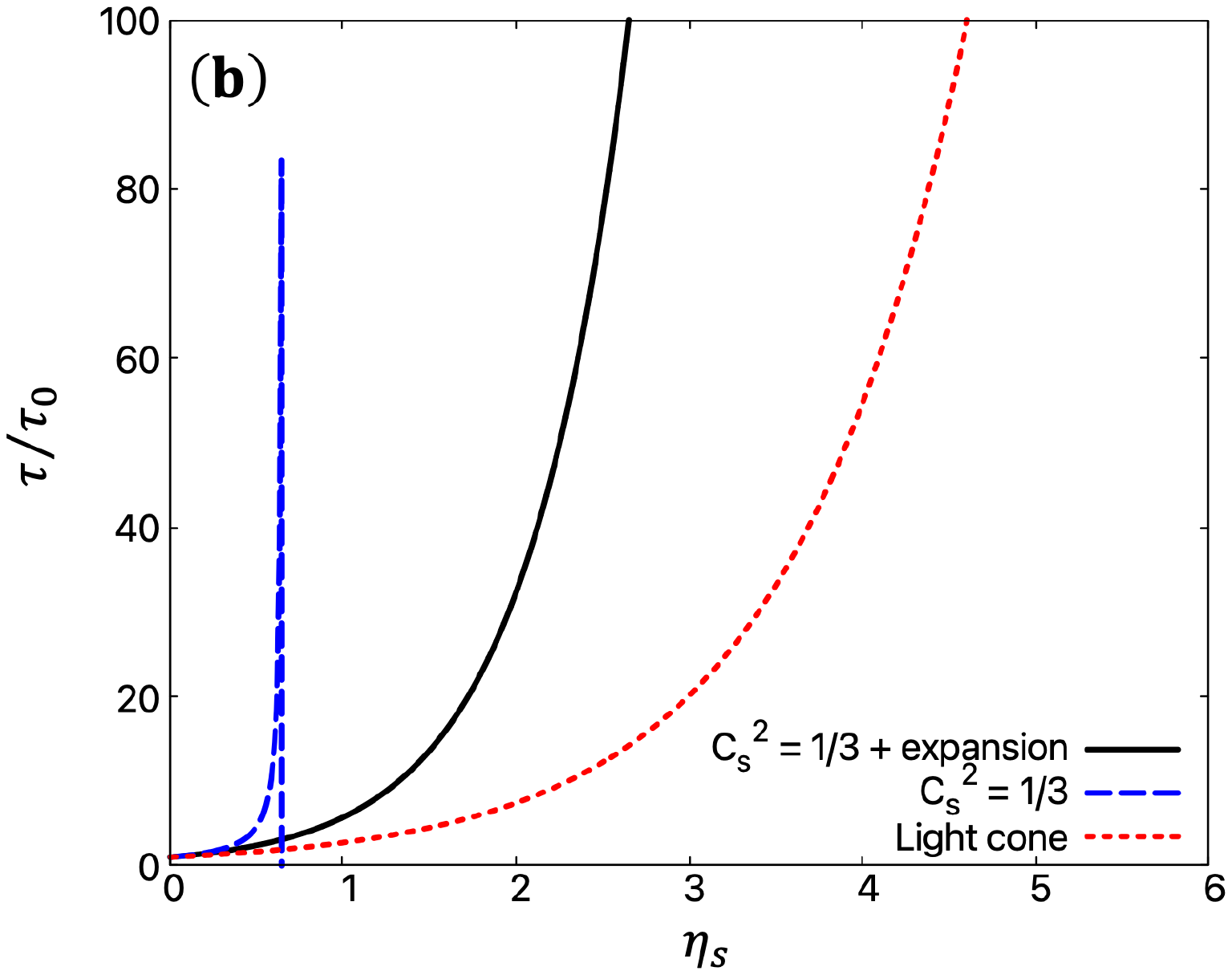}
    \end{minipage}
    \caption{Sound propagation in (a) the $t$-$z$ plane with an initial condition ($t_{0}=1 \;\text{fm}$ and $z_{0}=0 \;\text{fm}$) and in (b) the $\tau$-$\eta_{s}$ plane with an initial condition ($\tau_{0} = t_{0}$ and $\eta_{s0} = \eta_{s}(t_{0},0) = 0$).
    The black solid line, blue dashed line, and red dotted line denote propagation with tailwind, that without tailwind, and light cone, respectively. Sound velocity $c_{s}=1/\sqrt{3}$ is assumed.
    }
    \label{fig:sound}
\end{figure}

Here let us discuss the propagation of sound waves (or density fluctuations) in the $t$-$z$ plane and $\tau$-$\eta_{s}$ plane, 
and show that the behavior is completely different if the background medium expands.
One can define the effective sound velocity $c_{s}^{\prime}$, in which the tailwind from the expanding system is included, as
\begin{equation}\label{eq:propagation}
c_{s}^{\prime}(t,z) = \frac{dz(t)}{dt}=\tanh \left( \eta_{s}(t,z) + y_{s} \right) = \frac{c_{s}t+z(t)}{t+c_{s}z(t)},
\end{equation}
where $\eta_{s}(t,z) = \tanh^{-1}(z/t)$ and $y_{s} \equiv \tanh^{-1} c_{s}$, and $c_{s}$ is a constant sound velocity.
By performing a variable transformation $z(t)/t = v(t)$, one solves Eq.~(\ref{eq:propagation}) with an initial condition, $z(t=t_{0}) = z_{0}$, transforms the solution to $\tau$ and $\eta_{s}$, and obtains $\tau/\tau_{0}=e^{\Delta \eta_{s}/c_{s}}$, where $\Delta \eta_{s} \equiv |\eta_{s}(t,z) - \eta_{s0}|$.
Here $\tau_{0} = \sqrt{t_{0}^{2} - z_{0}^{2}}$ and $\eta_{s0} = \eta_{s}(t_{0},z_{0})$.
On the other hand, propagation of light is described as either $t-t_{0} = z-z_{0}$ in the $t$-$z$ plane or $\tau/\tau_{0}=e^{\Delta \eta_{s}}$ in the $\tau$-$\eta_{s}$ plane.
Figure \ref{fig:sound} shows the resultant propagation of sounds in (a) the $t$-$z$ plane and (b) the $\tau$-$\eta_{s}$ plane with sound velocity $c_{s}=1/\sqrt{3}$ and specific initial conditions.
In contrast to the case without tailwind, the slope of the propagation with tailwind approaches that of light cone asymptotically in $t$-$z$ plane. In other words, the propagation speed of sound approaches the speed of light in the long time limit even though the initial sound velocity is $c_{s}=1/\sqrt{3}$ due to the expansion of the system.
Moreover, propagation of sounds goes to infinity in $\eta_{s}$ space in this case.
On the other hand, there is a singularity at $\eta_{s} = y_{s} = \tanh^{-1}(1/\sqrt{3}) \simeq 0.66$ in the $\tau$-$\eta_{s}$ plane in the case without tailwind.
This is the so-called sound horizon.

\section{Two-particle correlation functions}\label{sec:B}
We obtain the event-averaged two-particle correlation function as a function of rapidity gap after inserting the explicit forms of distribution functions (\ref{eq:fideal}) and (\ref{eq:fvis}) into the following form:
\begin{align}\label{eq:two particle correlation 1}
\left< \frac{dN}{dY_{1}}\frac{dN}{dY_{2}} \right>
&=\left[\frac{d\tau A}{(2\pi)^{2}}\right]^{2} \int_{Y_{1} - \Delta\eta_{s1}^{\prime}}^{Y_{1} + \Delta\eta_{s1}^{\prime}} d\eta_{s1} \int_{Y_{2} - \Delta\eta_{s2}^{\prime}}^{Y_{2} + \Delta\eta_{s2}^{\prime}} d\eta_{s2}\nonumber\\
&\quad\times \int_{m}^{\infty}dm_{T1} \int_{m}^{\infty}dm_{T2} m_{T1}^{2} m_{T2}^{2} \nonumber\\
&\quad\times\cosh(Y_{1}-\eta_{s1}) \cosh(Y_{2}-\eta_{s2})\nonumber\\
&\quad\times \left< \left(f_{\rm{ideal,0}}^{(1)}f_{\rm{ideal,0}}^{(2)} + 2f_{\rm{ideal,0}}^{(1)}f_{\rm{vis,0}}^{(2)} + f_{\rm{vis,0}}^{(1)}f_{\rm{vis,0}}^{(2)}\right. \right.\nonumber\\
&\quad\times\left. \left.+ \delta f_{\rm{ideal}}^{(1)}\delta f_{\rm{ideal}}^{(2)} + 2\delta f_{\rm{ideal}}^{(1)}\delta f_{\rm{vis}}^{(2)} + \delta f_{\rm{vis}}^{(1)}\delta f_{\rm{vis}}^{(2)}  \right) \right>.
\end{align}
Here upper indices 1 and 2 in the distribution functions denote variables of particles 1 and 2, respectively.
Terms including single $\delta f_{\rm{ideal}}$ or $\delta f_{\rm{vis}}$ are neglected since they do not contribute to the final results due to $\left< \delta f \right> = 0$.
The next step is to integrate Eq.~(\ref{eq:two particle correlation 1}) over $m_{T1}$ and $m_{T2}$.
Since the resultant form of two-particle correlation functions is significantly complicated, we introduce new variables as follows to avoid the complexity:
\begin{align}
A_{1/2}(n) &\equiv \sum_{i=0}^{n} \frac{T_{0}^{i+1} P(n, i)m_{T1/2}^{n-i}}{\cosh^{i+1}(Y_{1/2}-\eta_{s1/2})},\\
B &\equiv \frac{1}{2s_{0}T_{0}^{3}},\\
C_{1/2} &\equiv \frac{1}{2} - \sinh^{2}(Y_{1/2}-\eta_{s1/2}),\\
D &\equiv \frac{1}{2}m^{2},\\
E &\equiv \frac{1}{2s_{0}T_{0}}\left(\frac{1}{c_{s}^{2}}+3\right),\\
F_{1/2} &\equiv 2\sinh(Y_{1/2}-\eta_{s1/2})\cosh(Y_{1/2}-\eta_{s1/2}),\\
G_{12} &\equiv \frac{\cosh(Y_{1}-\eta_{s1})\cosh(Y_{2}-\eta_{s2})}{T_{0}^{4}}\delta T(\eta_{s1}) \delta T(\eta_{s2})\nonumber\\
&\quad+ \frac{\sinh(Y_{1}-\eta_{s1})\cosh(Y_{2}-\eta_{s2})}{T_{0}^{3}}\delta y(\eta_{s1}) \delta T(\eta_{s2})\nonumber\\
&\quad+ \frac{\cosh(Y_{1}-\eta_{s1})\sinh(Y_{2}-\eta_{s2})}{T_{0}^{3}}\delta T(\eta_{s1}) \delta y(\eta_{s2})\nonumber\\
&\quad+ \frac{\sinh(Y_{1}-\eta_{s1})\sinh(Y_{2}-\eta_{s2})}{T_{0}^{2}}\delta y(\eta_{s1}) \delta y(\eta_{s2}).
\end{align}
Here the index ``$1/2$" denotes either particle 1 or particle 2.
A coefficient of the incomplete gamma function, $A_{1/2}(n)$,  is defined as
\begin{align}
&\int_{m}^{\infty} dm_{T1/2} m_{T1/2}^{n} e^{-\frac{m_{T1/2}\cosh(Y_{1/2}-\eta_{s1/2})}{T_{0}}}\nonumber\\
&= \sum_{i=0}^{n} \frac{T_{0}^{i+1} P(n,i)m_{T1/2}^{n-i}}{\cosh^{i+1}(Y_{1/2}-\eta_{s1/2})} e^{-\frac{m\cosh(Y_{1/2}-\eta_{s1/2})}{T_{0}}}\nonumber\\
&= A_{1/2}(n)e^{-\frac{m\cosh(Y_{1/2}-\eta_{s1/2})}{T_{0}}},
\end{align}
where the permutation is calculated as $P(n,i) = n!/(n-i)!$.
Using these variables, the resultant two-particle correlation leads to
\begin{widetext}
\begin{align}\label{eq:two particle correlation 2}
\left< \frac{dN}{dY_{1}}\frac{dN}{dY_{2}} \right>
&=\left[\frac{d\tau A}{(2\pi)^{2}}\right]^{2} \int_{Y_{1} - \Delta\eta_{s1}^{\prime}}^{Y_{1} + \Delta\eta_{s1}^{\prime}} d\eta_{s1} \int_{Y_{2} - \Delta\eta_{s2}^{\prime}}^{Y_{2} + \Delta\eta_{s2}^{\prime}} d\eta_{s2} \cosh(Y_{1}-\eta_{s1}) \cosh(Y_{2}-\eta_{s2})\nonumber\\
&\quad\times \left< \left( \mathrm{(I)} + 2\times\mathrm{(II)} + \mathrm{(III)} + \mathrm{(IV)} + 2\times\mathrm{(V)} + \mathrm{(VI)}  \right)\right>,\\
\mathrm{(I)} &= \int_{m}^{\infty}dm_{T1} \int_{m}^{\infty}dm_{T2} m_{T1}^{2} m_{T2}^{2} f_{\rm{ideal,0}}^{(1)}f_{\rm{ideal,0}}^{(2)} = A_{1}(2)A_{2}(2) e^{-\frac{m\left[ \cosh(Y_{1}-\eta_{s1})+ \cosh(Y_{2}-\eta_{s2}) \right]}{T_{0}}},\\
\mathrm{(II)} &= \int_{m}^{\infty}dm_{T1} \int_{m}^{\infty}dm_{T2} m_{T1}^{2} m_{T2}^{2} f_{\rm{ideal,0}}^{(1)}f_{\rm{vis,0}}^{(2)}
= \big[ A_{1}(2)A_{2}(4)BC_{2}\pi_{0}-A_{1}(2)A_{2}(2)BD\pi_{0} \big]
 e^{-\frac{m\left[ \cosh(Y_{1}-\eta_{s1})+ \cosh(Y_{2}-\eta_{s2}) \right]}{T_{0}}},\\
\mathrm{(III)} &= \int_{m}^{\infty}dm_{T1} \int_{m}^{\infty}dm_{T2} m_{T1}^{2} m_{T2}^{2} f_{\rm{vis,0}}^{(1)}f_{\rm{vis,0}}^{(2)}
= \big[ A_{1}(4)A_{2}(4)B^{2}C_{1}C_{2}\pi_{0}^{2} - A_{1}(4)A_{2}(2)B^{2}C_{1}D\pi_{0}^{2}\nonumber\\
&\quad-A_{1}(2)A_{2}(4)B^{2}C_{2}D\pi_{0}^{2} + A_{1}(2)A_{2}(2)B^{2}D^{2}\pi_{0}^{2} \big] e^{-\frac{m\left[ \cosh(Y_{1}-\eta_{s1})+ \cosh(Y_{2}-\eta_{s2}) \right] }{T_{0}}},\\
\mathrm{(IV)} &= \int_{m}^{\infty}dm_{T1} \int_{m}^{\infty}dm_{T2} m_{T1}^{2} m_{T2}^{2} \delta f_{\rm{ideal}}^{(1)}\delta f_{\rm{ideal}}^{(2)} = A_{1}(3)A_{2}(3)G_{12} e^{-\frac{m\left[ \cosh(Y_{1}-\eta_{s1})+ \cosh(Y_{2}-\eta_{s2}) \right]}{T_{0}}},\\
\mathrm{(V)} &= \int_{m}^{\infty}dm_{T1} \int_{m}^{\infty}dm_{T2} m_{T1}^{2} m_{T2}^{2} \delta f_{\rm{ideal}}^{(1)}\delta f_{\rm{vis}}^{(2)}\nonumber\\
&= \Bigg\{ A_{1}(3)A_{2}(5)G_{12}BC_{2}\pi_{0} - A_{1}(3)A_{2}(3)G_{12}BD\pi_{0}\nonumber\\
&\quad-A_{1}(3)A_{2}(4)EC_{2}\pi_{0}\Big[ \frac{\sinh(Y_{1}-\eta_{s1})}{T_{0}}\delta y(\eta_{s1})\delta T(\eta_{s2}) + \frac{\cosh(Y_{1}-\eta_{s1})}{T_{0}^{2}}\delta T(\eta_{s1})\delta T(\eta_{s2}) \Big]\nonumber\\
&\quad+A_{1}(3)A_{2}(4)BF_{2}\pi_{0}\Big[ \frac{\sinh(Y_{1}-\eta_{s1})}{T_{0}}\delta y(\eta_{s1})\delta y(\eta_{s2})
+ \frac{\cosh(Y_{1}-\eta_{s1})}{T_{0}^{2}}\delta T(\eta_{s1})\delta y(\eta_{s2}) \Big]\nonumber\\
&\quad+A_{1}(3)A_{2}(4)BC_{2}\Big[ \frac{\sinh(Y_{1}-\eta_{s1})}{T_{0}}\delta y(\eta_{s1})\delta \pi(\eta_{s2})
+ \frac{\cosh(Y_{1}-\eta_{s1})}{T_{0}^{2}}\delta T(\eta_{s1})\delta \pi(\eta_{s2}) \Big]\nonumber\\
&\quad+A_{1}(3)A_{2}(2)ED\pi_{0}\Big[ \frac{\sinh(Y_{1}-\eta_{s1})}{T_{0}}\delta y(\eta_{s1})\delta T(\eta_{s2})+ \frac{\cosh(Y_{1}-\eta_{s1})}{T_{0}^{2}}\delta T(\eta_{s1})\delta T(\eta_{s2}) \Big]\nonumber\\
&\quad-A_{1}(3)A_{2}(2)BD\Big[ \frac{\sinh(Y_{1}-\eta_{s1})}{T_{0}}\delta y(\eta_{s1})\delta \pi(\eta_{s2})+ \frac{\cosh(Y_{1}-\eta_{s1})}{T_{0}^{2}}\delta T(\eta_{s1})\delta \pi(\eta_{s2}) \Big] \Bigg\}\nonumber\\
&\quad\times e^{-\frac{m\left[ \cosh(Y_{1}-\eta_{s1})+ \cosh(Y_{2}-\eta_{s2}) \right] }{T_{0}}},\\
\mathrm{(VI)} &= \int_{m}^{\infty}dm_{T1} \int_{m}^{\infty}dm_{T2} m_{T1}^{2} m_{T2}^{2} \delta f_{\rm{vis}}^{(1)}\delta f_{\rm{vis}}^{(2)}\nonumber\\
&= \Bigg( \Big\{ A_{1}(5)A_{2}(5)G_{12}B^{2}C_{1}C_{2}\pi_{0}^{2} - A_{1}(5)A_{2}(3)G_{12}B^{2}C_{1}D\pi_{0}^{2}\nonumber\\
&\quad- A_{1}(3)A_{2}(5)G_{12}B^{2}C_{2}D\pi_{0}^{2} + A_{1}(3)A_{2}(3)G_{12}B^{2}D^{2}\pi_{0}^{2}+ \Big[ A_{1}(4)A_{2}(4)E^{2}C_{1}C_{2}\pi_{0}^{2} - A_{1}(4)A_{2}(2)E^{2}C_{1}D\pi_{0}^{2}\nonumber\\
&\quad- A_{1}(2)A_{2}(4)E^{2}C_{2}D\pi_{0}^{2} + A_{1}(2)A_{2}(2)E^{2}D^{2}\pi_{0}^{2} \Big] \delta T(\eta_{s1})\delta T(\eta_{s2})
+ A_{1}(4)A_{2}(4)B^{2}F_{1}F_{2}\pi_{0}^{2} \delta y(\eta_{s1})\delta y(\eta_{s2})\nonumber\\
&\quad+ \Big[ A_{1}(4)A_{2}(4)B^{2}C_{1}C_{2} - A_{1}(4)A_{2}(2)B^{2}C_{1}D
- A_{1}(2)A_{2}(4)B^{2}C_{2}D + A_{1}(2)A_{2}(2)B^{2}D^{2} \Big]
\delta \pi(\eta_{s1})\delta \pi(\eta_{s2}) \Big\} \nonumber\\
&\quad+ 2\Big\{ \Big[ -A_{1}(5)A_{2}(4)BEC_{1}C_{2}\pi_{0}^{2} + A_{1}(3)A_{2}(4)BEDC_{2}\pi_{0}^{2}
+ A_{1}(5)A_{2}(2)BEC_{1}D\pi_{0}^{2} - A_{1}(3)A_{2}(2)BED^{2}\pi_{0}^{2} \Big] \nonumber\\
\end{align}
\begin{align}
&\times \Big[ \frac{\sinh(Y_{1}-\eta_{s1})}{T_{0}}\delta y(\eta_{s1})\delta T(\eta_{s2})+ \frac{\cosh(Y_{1}-\eta_{s1})}{T_{0}^{2}}\delta T(\eta_{s1})\delta T(\eta_{s2}) \Big]
+ \Big[ A_{1}(5)A_{2}(4)B^{2}C_{1}F_{2}\pi_{0}^{2}\nonumber\\
&- A_{1}(3)A_{2}(4)B^{2}DF_{2}\pi_{0}^{2} \Big] \Big[ \frac{\sinh(Y_{1}-\eta_{s1})}{T_{0}}\delta y(\eta_{s1})\delta y(\eta_{s2})
+ \frac{\cosh(Y_{1}-\eta_{s1})}{T_{0}^{2}}\delta T(\eta_{s1})\delta y(\eta_{s2}) \Big]\nonumber\\
&+ \Big[ A_{1}(5)A_{2}(4)B^{2}C_{1}C_{2}\pi_{0} - A_{1}(3)A_{2}(4)B^{2}DC_{2}\pi_{0}
- A_{1}(5)A_{2}(2)B^{2}C_{1}D\pi_{0} + A_{1}(3)A_{2}(2)B^{2}D^{2}\pi_{0} \Big] \nonumber\\
&\times \Big[ \frac{\sinh(Y_{1}-\eta_{s1})}{T_{0}}\delta y(\eta_{s1})\delta \pi(\eta_{s2})
+ \frac{\cosh(Y_{1}-\eta_{s1})}{T_{0}^{2}}\delta T(\eta_{s1})\delta \pi(\eta_{s2}) \Big]\nonumber\\
&+ \Big[ -A_{1}(4)A_{2}(4)BEC_{1}F_{2}\pi_{0}^{2} + A_{1}(2)A_{2}(4)BEDF_{2}\pi_{0}^{2} \Big]
\delta T(\eta_{s1})\delta y(\eta_{s2})\nonumber\\
&+ \Big[ -A_{1}(4)A_{2}(4)BEC_{1}C_{2}\pi_{0} + A_{1}(4)A_{2}(2)BEC_{1}D\pi_{0}
+ A_{1}(2)A_{2}(4)BEC_{2}D\pi_{0} - A_{1}(2)A_{2}(2)BED^{2}\pi_{0} \Big] \nonumber\\
& \times \delta T(\eta_{s1})\delta \pi(\eta_{s2})
+ \Big[ A_{1}(4)A_{2}(4)B^{2}F_{1}C_{2}\pi_{0} - A_{1}(4)A_{2}(2)B^{2}F_{1}D\pi_{0} \Big] \delta y(\eta_{s1})\delta \pi(\eta_{s2}) \Big\} \Bigg) e^{-\frac{m\left[ \cosh(Y_{1}-\eta_{s1})+ \cosh(Y_{2}-\eta_{s2}) \right] }{T_{0}}}.
\end{align}
\end{widetext}
The terms \rm{(IV)}, \rm{(V)}, and \rm{(VI)} play a crucial role in reflecting the correlations of thermodynamic variables in the two-particle correlation functions.
Performing $\eta_{s1}$ and $\eta_{s2}$ integration of (\ref{eq:two particle correlation 2}) numerically, we finally obtain the two-particle correlation as a function of rapidity gap, $\Delta Y \equiv |Y_{1}-Y_{2}|$.

In Eq.~(\ref{eq:two particle correlation 2}), we considered up to the second order in perturbation (fluctuation) and up to the second order in viscous correction.
Orders of each term are summarized in Table \ref{tb:order}.
Strictly speaking, the second-order perturbative terms from the one-particle distribution or the second order viscous corrections should have been considered from an order counting point of view in the form of two-particle correlation (\ref{eq:two particle correlation 2}), e.g., $f_{\rm{ideal,0}}\delta^{2}f_{\rm{vis}}$.
However, we neglected such terms since we expanded the distribution only up to the first order of perturbation and viscous correction as Eq.~(\ref{eq:expansion}).
\begin{table}[htbp]
\centering
\caption{Orders of each term.}
\begin{tabular}{lcc} 
\hline
\hline
Term & Perturbation & Viscous correction\\ \hline
$f_{\rm{ideal,0}}f_{\rm{ideal,0}}$ & 0 & 0\\
$f_{\rm{ideal,0}}f_{\rm{vis,0}}$ & 0 & 1\\
$f_{\rm{vis,0}}f_{\rm{vis,0}}$ & 0 & 2\\
$\delta f_{\rm{ideal}}\delta f_{\rm{ideal}}$ & 2 & 0\\
$\delta f_{\rm{ideal}}\delta f_{\rm{vis}}$ & 2 & 1\\
$\delta f_{\rm{vis}}\delta f_{\rm{vis}}$ & 2 & 2\\\hline
\end{tabular}
\label{tb:order}
\end{table}
\bibliography{reference.bib}

\end{document}